\documentclass[nobibnotes,preprintnumbers,aps,prl,superscriptaddress,showpacs,twocolumn,twoside,sort&compress]{revtex4}				
\usepackage{graphicx}																												
\usepackage{titlesec}	
\usepackage{multirow}	
\usepackage{gensymb}																				\usepackage{ulem}	
\usepackage{color,soul}
\usepackage[colorlinks=true,linkcolor=blue,citecolor=red,urlcolor=blue]{hyperref}													

\begin{document}																													
%
																												

\newcommand{\kvec}{\mbox{{\scriptsize {\bf k}}}}
\newcommand{\lvec}{\mbox{{\scriptsize {\bf l}}}}
\newcommand{\qvec}{\mbox{{\scriptsize {\bf q}}}}

\def\eq#1{(\ref{#1})}
\def\fig#1{\hspace{1mm}Fig. \ref{#1}}
\def\tab#1{\hspace{1mm}Table \ref{#1}}
\title{Isotope effect in superconducting lanthanum hydride under high compression}

\author{Artur P. Durajski} \email{adurajski@wip.pcz.pl}
\affiliation{Institute of Physics, Cz{\c{e}}stochowa University of Technology, Ave. Armii Krajowej 19, 42-200 Cz{\c{e}}stochowa, Poland}
\author{Rados{\l}aw Szcz{\c{e}}{\'s}niak}
\affiliation{Institute of Physics, Cz{\c{e}}stochowa University of Technology, Ave. Armii Krajowej 19, 42-200 Cz{\c{e}}stochowa, Poland}
\author{Yinwei Li}
\affiliation{Laboratory of Quantum Materials Design and Application, School of Physics and Electronic Engineering, Jiangsu Normal University, Xuzhou 221116, China}
\author{Chongze Wang}
\affiliation{Department of Physics, Research Institute for Natural Science,
and HYU-HPSTAR-CIS High Pressure Research Center, Hanyang University,
222 Wangsimni-ro, Seongdong-Ku, Seoul 04763, Republic of Korea}
\author{Jun-Hyung Cho} \email{chojh@hanyang.ac.kr}
\affiliation{Department of Physics, Research Institute for Natural Science,
and HYU-HPSTAR-CIS High Pressure Research Center, Hanyang University,
222 Wangsimni-ro, Seongdong-Ku, Seoul 04763, Republic of Korea}

\date{\today}

\begin{abstract}
Recently, the discovery of room-temperature superconductivity (SC) was experimentally realized in the fcc phase of LaH$_{10}$ under megabar pressure. Specifically, the isotope effect of $T_{\rm c}$ was measured by the replacement of hydrogen (H) with deuterium (D), demonstrating a driving role of phonons in the observed room-temperature SC. Herein, based on the first-principles calculations within the harmonic approximation, we reveal that (i) the identical electron-phonon coupling constants of fcc LaH$_{10}$ and LaD$_{10}$ decrease monotonously with increasing pressure and (ii) the isotope effect of $T_{\rm c}$ is nearly proportional to $M^{-{\alpha}}$ ($M$: ionic mass) with ${\alpha}$ ${\approx}$ 0.465, irrespective of pressure. The predicted value of ${\alpha}$ agrees well with the experimental one (${\alpha}=0.46$) measured at around 150 GPa. Thus, our findings provide a theoretical confirmation of the conventional electron-phonon coupling mechanism in a newly discovered room-temperature superconductor of compressed LaH$_{10}$.
\end{abstract}
\maketitle
%
\section{I. Introduction}

Ever since the first discovery of superconductivity (SC) in 1911~\cite{onnes}, the realization of room-temperature SC has been the holy grail of physics. Based on the Bardeen-Cooper-Schrieffer (BCS) theory~\cite{Bardeen1957A}, Neil Ashcroft~\cite{Ashcroft1968} proposed that the metallic hydrogen formed under high pressures over ${\sim}400$ GPa~\cite{MetalicH1, MetalicH2} would be an ideal candidate for room-temperature SC. In order to achieve the metallization of hydrogen lattices at relatively lower pressures attainable in the diamond anvil cells~\cite{diamondanvil1, diamondanvil2}, many binary hydrides have been theoretically searched~\cite{Wang2012-CaH6, Duan2014A, Feng2015-MgH6, rare-earth-hydride1, rare-earth-hydride2, Hydride1, Hydride2, LiY2015, PhysRevB.93.020103, QuanYundi, PhysRevB.98.100102}. Due to the \textit{chemical pre-compression}, such hydrides can have high superconducting transition temperatures $T_{\rm c}$ at relatively lower pressures. For instance, theoretically predicted \cite{Li2014A, Duan2014A} and then experimentally realized cubic structure of hydrogen sulfide H$_3$S with a crystalline symmetry of the space group Im$\overline{3}$m exhibits SC with a $T_c$ of ${\sim}203$ K at a pressure of $155$ GPa \cite{Drozdov2015A, Einaga2016A}. Furthermore, the pairing mechanism of SC in H$_3$S is associated with conventional electron-phonon interaction because the measurements of $T_c$ for H$_3$S and its deuterium counterpart D$_3$S showed a strong isotope effect: i.e., $T_c$ of D$_3$S shifts towards a lower temperature \cite{Drozdov2015A}.

Recently, two experimental groups synthesized a lanthanum hydride LaH$_{10}$ with a clathratelike structure at megabar pressures and measured a $T_{\rm c}$ between 250 and 260 K at a pressure of ${\sim}$170 GPa~\cite{ExpLaH10-PRL, Drozdov2019}. This record of $T_{\rm c}$ is the highest among so far experimentally available superconducting materials~\cite{Drozdov2015A, FeH5, CeH9, YH6}, which will bring a new era of high-$T_{\rm c}$ SC.
The x-ray diffraction and optical studies of lanthanum hydrides showed the existence of fcc lattice at ${\sim}170$ GPa upon heating to ${\sim}1000$ K~\cite{LaH10-angew}, consistent with the earlier predicted metallic fcc phase of LaH$_{10}$ having cages of $32$ H atoms surrounding a La atom~\cite{rare-earth-hydride1, rare-earth-hydride2, Kruglov, LiuLiangliang}. To reveal the underlying mechanism of the observed room-temperature SC in fcc LaH$_{10}$, a pronounced isotope shift was measured by the substitution of deuterium for hydrogen, thereby providing a direct evidence of the conventional phonon-mediated pairing mechanism~\cite{Drozdov2019}. Based on the relation of $T_{\rm c}$ ${\propto}$ $M^{-\alpha}$ \cite{Isotopecoeff}, where ${M}$ is ionic mass, the isotope coefficient $\alpha$ can be estimated. According to the experimental data of $T_{\rm c}=249$ (180) K for fcc LaH$_{10}$ (fcc LaD$_{10}$) at a pressure of ${\sim}$150 GPa~\cite{Drozdov2019}, the estimated value of $\alpha$ amounts to 0.46, close to that (${\alpha \approx 0.5}$) obtained from the BCS theory~\cite{Bardeen1957A}.

In this paper, using first-principles density-functional theory (DFT) calculations within the harmonic approximation, we investigate the isotope effect of fcc LaH$_{10}$ and fcc LaD$_{10}$ as a function of pressure. We find that the electron-phonon coupling (EPC) constant $\lambda$, which is invariant with respect to the H isotope substitution, decreases with increasing pressure as 2.38, 1.82, and 1.54 at 250, 300, and 350 GPa, respectively. By solving the Eliashberg equations, $T_{\rm c}$ of LaH$_{10}$ (LaD$_{10}$) decreases almost linearly as 234 (169), 214 (155), 195 (142) K at 250, 300, and 350 GPa, respectively. As a result, the isotope coefficient is estimated to be ${\sim}$0.465, irrespective of pressure. This estimated value is in good agreement with the experimental measurement \cite{Drozdov2019} of ${\alpha}=0.46$ at around 150 GPa. Therefore, our first-principles calculations strongly support a conventional electron-phonon coupling mechanism in the observed room-temperature SC of fcc LaH$_{10}$.

\section{II. Theoretical model and computational methods}

All the numerical calculations were performed using the DFT as implemented in the Quantum Espresso software package \cite{QE, QE2}.
The ultrasoft pseudopotentials were used for all atoms and the exchange-correlation energy was described by the Perde-Burke-Ernzerhof (PBE) functional based on the generalized gradient approximation (GGA).
To obtain the optimized geometry of lanthanum hydride systems at high pressure, the atomic positions and cell vectors were fully relaxed by using the Broyden-Fletcher-Goldfarb-Shanno (BFGS) quasi-Newton algorithm \cite{bfgs} up to the convergence criteria of less than $10^{-10}$ Ry and $10^{-8}$ kbar for total energy and pressure, respectively.
On the basis of convergence tests, the kinetic energy cutoff for the wave functions and charge density were taken as $80$ Ry and $1000$ Ry, respectively.
For self-consistent calculations, the $24\times 24\times 24$ k-points meshes were used.
Phonon spectra and electron-phonon interactions were calculated using the density functional perturbation theory \cite{BaroniDFPT}. Here, the first Brillouin zone was sampled using the $6\times 6\times 6$ q-point meshes and the denser $36\times 36\times 36$ k-point meshes, respectively.

The thermodynamic properties of superconducting states in compressed LaH$_{10}$ and LaD$_{10}$ are obtained by solving the isotropic Eliashberg equations with the superconducting order parameter function $\varphi_{n}=\varphi\left(i\omega_{n}\right)$ and the electron mass renormalization function $Z_{n}= Z\left(i\omega_{n}\right)$. The set of isotropic Eliashberg equations defined on the imaginary-frequency axis gives the following forms \cite{Eliashberg1960A, Marsiglio1988A}:
\begin{equation}
\label{r1}
\varphi_{n}=\frac{\pi}{\beta}\sum_{m=-M_f}^{M_f}
\frac{\lambda_{n,m}-\mu^{\star}\theta\left(\omega_{c}-|\omega_{m}|\right)}
{\sqrt{\omega_m^2Z^{2}_{m}+\varphi^{2}_{m}}}\varphi_{m},
\end{equation}
\vspace{-0.2cm}
\begin{equation}
\label{r2}
Z_{n}=1+\frac{1}{\omega_{n}}\frac{\pi}{\beta}\sum_{m=-M_f}^{M_f}
\frac{\lambda_{n,m}}{\sqrt{\omega_m^2Z^{2}_{m}+\varphi^{2}_{m}}}
\omega_{m}Z_{m},
\end{equation}
where the electron-phonon interaction pairing kernel is given by:
\begin{equation}
\label{r3}
\lambda_{n,m}= 2\int_0^{\infty}d\omega\frac{\omega}
{\left(\omega_n-\omega_m\right)^2+\omega ^2}\alpha^{2}F\left(\omega\right).
\end{equation}
Hence, the superconducting order parameter is defined by the ratio $\Delta_n=\varphi_{n}/Z_n$. The effective screened Coulomb repulsion constant $\mu^{\star}$ was chosen in the range of $0.1-0.2$, which can be adjusted with the comparison of the measured $T_c$ \cite{Durajski-50-2016, DominZemla}.
The Heaviside step function $\theta$ is determined by a frequency cutoff $\omega_c = 3$ eV , which is typically ten times larger than the maximum phonon frequency.
The value of $\beta$ is given by $\beta=1/k_BT$, where $k_B$ is the Boltzmann constant.
The Eliashberg spectral function $\alpha^{2}F(\omega)$, which is the main input element to the Eliashberg equations, is defined as:
\begin{equation}
\alpha^2F(\omega) = {1\over 2\pi N(\varepsilon_F)}\sum_{{\bf q}\nu}
                    \delta(\omega-\omega_{{\bf q}\nu})
                    {\gamma_{{\bf q}\nu}\over\hbar\omega_{{\bf q}\nu}}
\end{equation}
with
\begin{eqnarray}
\gamma_{{\bf q}\nu} &=& 2\pi\omega_{{\bf q}\nu} \sum_{ij}
                \int {d^3k\over \Omega_{BZ}}  |g_{{\bf q}\nu}({\bf k},i,j)|^2
                    \delta(\epsilon_{{\bf q},i} - \epsilon_F) \\\nonumber  &\times& \delta(\epsilon_{{\bf k}+{\bf q},j} - \epsilon_F),
\end{eqnarray}
where $N(\varepsilon_F)$, $\gamma_{{\bf q}\nu}$, and $g_{{\bf q}\nu}({\bf k},i,j)$ are the density of states at the Fermi energy $\varepsilon_F$, the phonon linewidth, and the electron-phonon matrix element, respectively.
The integrated EPC constant $\lambda(\omega)$ is obtained by the integration of ${\alpha^{2} F(\omega)}$:
\begin{equation}
\lambda(\omega)=2\int_{0}^{\omega} d\omega'{\alpha^2 F(\omega')}/{\omega'} ,
\end{equation}
where the total EPC constant is calculated as $\lambda(\omega\rightarrow\infty)$.

The Eliashberg equations are solved iteratively in a self-consistent way with a maximal error of $10^{-10}$ between two successive iterations. The convergence and precision are controlled by using the sufficiently high number ($M_f=1100$) of Matsubara frequencies: $\omega_{n}=\left(\pi/\beta\right)\left(2n-1\right)$, where $n=0,\pm 1,\pm 2,\dots,\pm M_f$.
For further details about the implementation and derivation of the methods adopted herein, see the references \cite{szczesniak2006A, Durajski-64-2018, Wiendlocha, PhysRevB.100.094505}.

\section{III. Results and discussion}

We first determine the geometry of Fm$\overline{3}$m LaH$_{10}$ phase in the pressure range of $250-350$ GPa using the first-principles DFT calculations.
Here, we consider three different pressures $250$, $300$, and $350$ GPa, because the experimentally observed fcc LaH$_{10}$ phase becomes unstable at lower pressures below $220$ GPa (see Fig. S1 in the Supplemental Material \cite{SM}).
The calculated pressure-volume data show the nearly linear reduction of volume with a slope of $\rm{d}V/\rm{d}p=-0.029$ $\rm{\AA}^3/\rm{GPa}$, where we obtain $28.484$, $26.891$, and $25.564$ $\rm{\AA}^3$ at $250$, $300$, and $350$ GPa, respectively.
Figure 1(a) compares the calculated electronic band structures of fcc LaH$_{10}$ at $250$, $300$, and $350$ GPa, whereas their corresponding densities of states around $\varepsilon_F$ are displayed in Fig. 1(b).
It is seen that the dispersions of the bands around $\varepsilon_F$ change very little with respect to pressure.
Consequently, the pressure dependence of the density of states (DOS) at $\varepsilon_F$ is minor, compared to those at the energy regions away from the $\varepsilon_F$.
The present results of electronic band structure and DOS with respect to pressure agree well with those obtained using the Vienna $ab$ $initio$ simulation package with the projector augmented-wave method \cite{chongze-prb}.
As shown in Fig. 1(b), the DOS at $\varepsilon_F$ reaches $0.83-0.86$ states/eV in the range of $250-350$ GPa, which are comparable with that ($0.63-0.90$ states/eV) of compressed H$_3$S having a $T_c$ of ${\sim}203$ K at 155 GPa \cite{BianconiSciRep, PhysRevB.93.104526}.
It is thus likely that both compressed LaH$_{10}$ and H$_3$S with high DOS at $\varepsilon_F$ would be equally expected to have high-$T_c$ SC, as observed by experiments \cite{Drozdov2015A, Einaga2016A, ExpLaH10-PRL, Drozdov2019}.
We note that the electronic band structure of fcc LaD$_{10}$ is identical to that of fcc LaH$_{10}$ because of the same Kohn-Sham effective potentials in both systems.

\begin{figure}[!h]
\includegraphics[width=0.9\columnwidth]{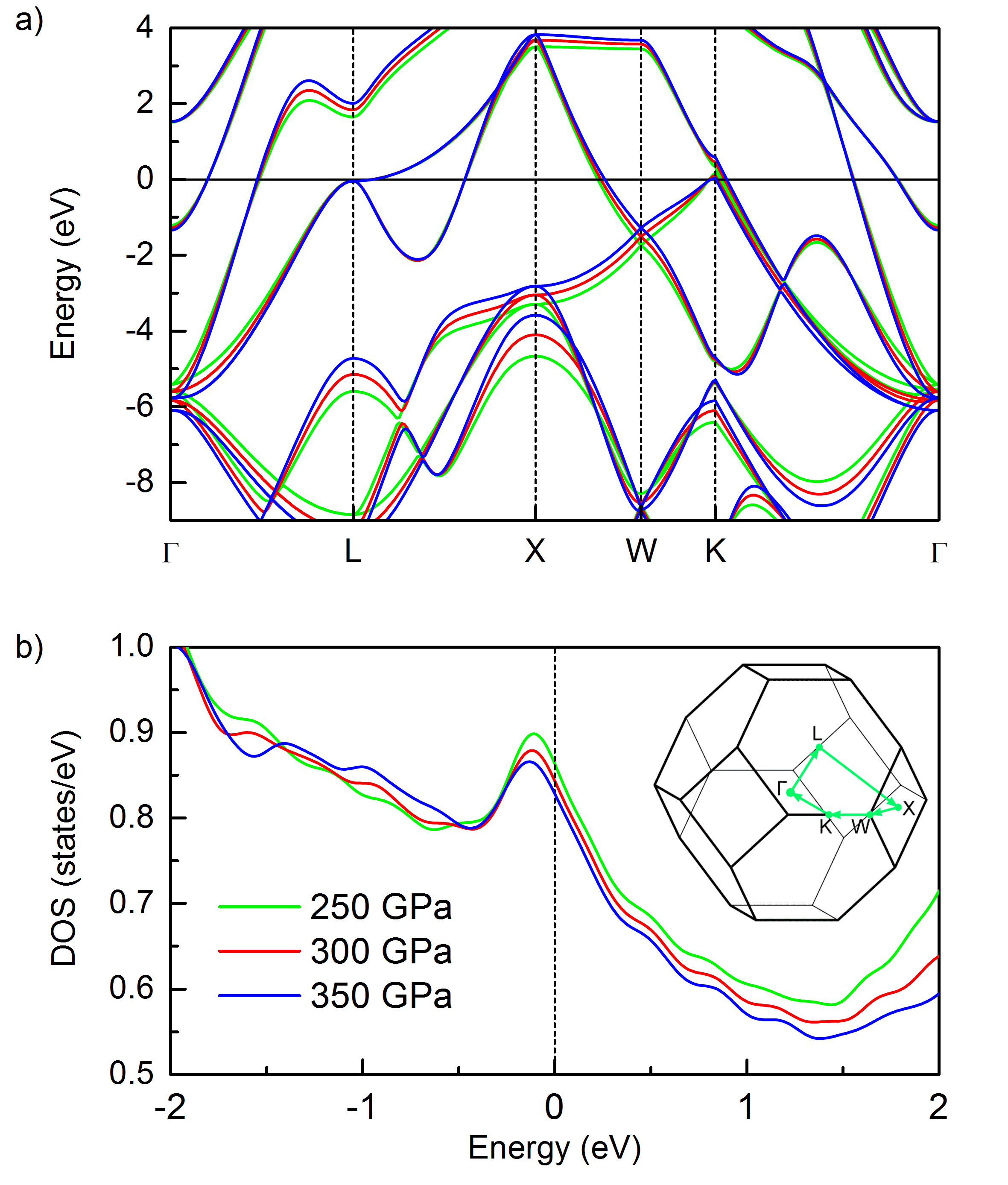}
\caption{Calculated electronic band structures of fcc LaH$_{10}$ at 250, 300, and 350 GPa: (a) The electronic dispersion along the high-symmetry lines in the Brillouin zone and (b) the total DOS per formula unit. The Fermi level $\epsilon_F$ is set to zero. The inset of (b) shows the Brillouin zone for cubic (fcc) clathrate-type structure with special k-point paths.}
\end{figure}
\begin{figure}[!h]
\includegraphics[width=1\columnwidth]{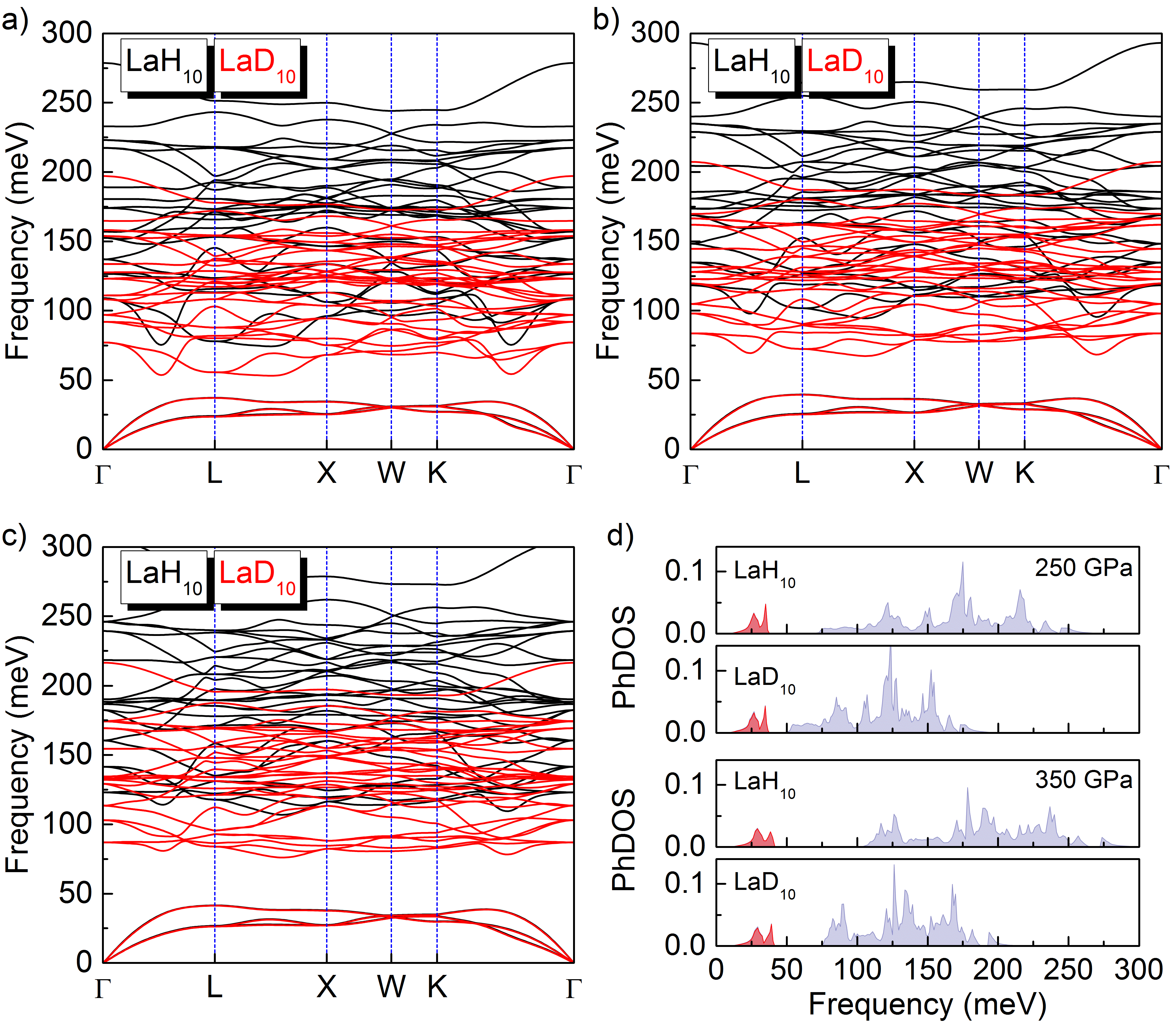}
\caption{Calculated phonon dispersions of fcc LaH$_{10}$ and fcc LaD$_{10}$ at (a) 250 GPa, (b) 300 GPa, and (c) 350 GPa. (d) Projected phonon density of states (PhDOS) onto La (red-color region) and H/D (gray-color region) atoms at 250 and 350 GPa.}
\label{f2}
\end{figure}

Figures 2(a), 2(b), and 2(c) display the calculated phonon dispersions of fcc LaH$_{10}$ at 250, 300, and 350 GPa, respectively, together with the overlap of the corresponding results for fcc LaD$_{10}$.
For each structure, there are no imaginary phonon frequencies, indicating that fcc LaH$_{10}$ and LaD$_{10}$ are dynamically stable in the pressure range between $250$ and $350$ GPa.
In Fig. 2(d), we plot the phonon DOS projected onto La and H atoms at $250$ and $350$ GPa.
We find that the acoustic phonon modes with lower frequencies below ${\sim}45$ meV arise from La atoms, which are well separated from the optical phonon modes of H or D atoms.
It is noticeable that the D-derived optical phonon modes shift towards lower frequencies, relative to the corresponding H-derived modes.
For instances, at $250$, $300$, and $350$ GPa, the lowest optical modes at the $\Gamma$ point shift from $109.44$, $118.36$, and $123.12$ meV in LaH$_{10}$ to $77.52$, $83.44$, and $86.93$ meV in LaD$_{10}$, respectively.
Thus, we can say that the frequencies of the H- and D-derived optical phonon modes are nearly inversely proportional to the square root of ionic mass, as discussed below.

\begin{figure}[!h]
\includegraphics[width=1\columnwidth]{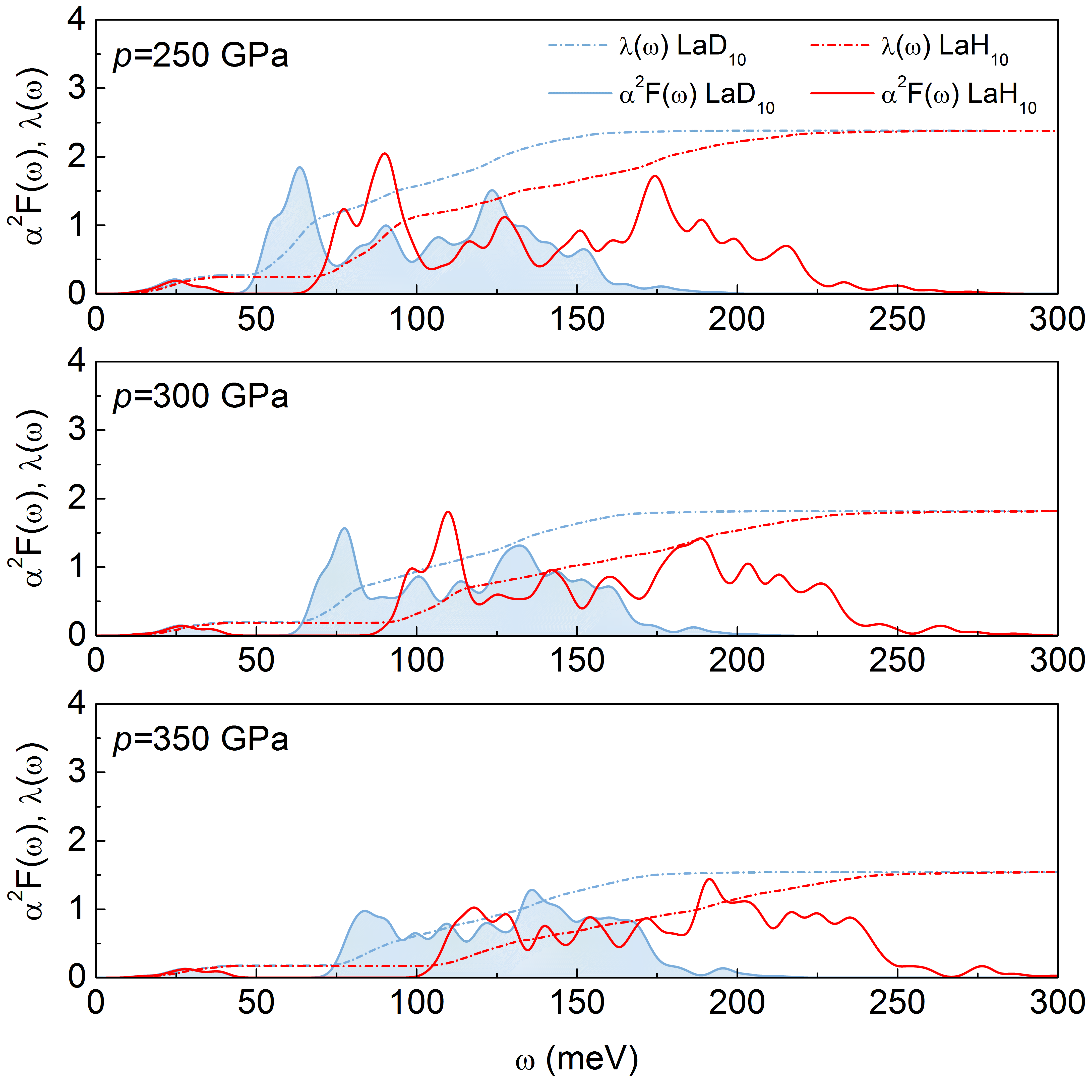}
\caption{Calculated Eliashberg spectral functions $\alpha ^2F(\omega)$ of fcc LaH$_{10}$ and LaD$_{10}$ at 250, 300, and 350 GPa, together with the corresponding integrated EPC constant $\lambda(\omega)$.}
\label{f3}
\end{figure}

Figure 3 shows the results of Eliashberg function $\alpha ^2F(\omega)$ and integrated EPC constant $\lambda(\omega)$ as a function of phonon frequency, calculated at $250$, $300$, and $350$ GPa.
We find that the La-derived acoustical phonon modes contribute to only about $11\%$ of the total $\lambda$.
This result indicates that H atoms play a dominant role in contributing to electron-phonon coupling, which in turn gives rise to a room-temperature SC in fcc LaH$_{10}$.
The calculated total $\lambda$ values of LaH$_{10}$ and LaD$_{10}$ are equally as 2.38, 1.82, and 1.54 at 250, 300, and 350 GPa, respectively.
Interestingly, as pressure increases, the total $\lambda$ decreases monotonously, consistent with the decrease of $T_c$ measured by a recent experiment of fcc LaH$_{10}$ \cite{Drozdov2019}.
Because of such large EPC constants larger than 1, we adopt the Eliashberg theory to understand the properties of superconducting state through conventional EPC mechanism.

We calculate the temperature dependence of the superconducting energy gap ${\Delta}$ by solving the Eliashberg equations on the imaginary- and real-frequency axes \cite{Marsiglio1988A}.
Figure 4 shows the calculated ${\Delta}$ $vs$ temperature curves at 250, 300, and 350 GPa, together with the real and imaginary parts of $\Delta\left(\omega,T=0\right)$ as a function of frequency.
Based on these data, we estimate $T_c$, $\Delta\left(0\right)$, and the universal dimensionless ratio $2\Delta\left(0\right)/k_{B}T_{c}$, all of which are influenced by pressure and Coulomb pseudopotential ${\mu}^{\star}$ \cite{Gonczarek2}. The results are listed in \tab{t1}. We find that at 250 GPa, $T_c$ is estimated as high as $269$ and $191$ K for LaH$_{10}$ and LaD$_{10}$, respectively.
Further, $T_c$ is found to decrease almost linearly with increasing pressure, reaching $237$ K for LaH$_{10}$ and $170$ K for LaD$_{10}$ at $350$ GPa.
As shown in \tab{t1}, the larger value of $\lambda$, the higher is the $2\Delta\left(0\right)/k_{B}T_{C}$ ratio.
Moreover, $2\Delta\left(0\right)/k_{B}T_{C}$ considerably exceeds the universal value of $3.53$ predicted by the BCS theory \cite{Bardeen1957A}, which reflects the strong EPC and retardation effects in fcc LaH$_{10}$ and LaD$_{10}$.

\begin{table*}[]
\centering
\caption{Calculated EPC constant $\lambda$, $T_c$, $\Delta(0)$, and dimensionless ratio $2\Delta\left(0\right)/k_{B}T_{C}$ of fcc LaH$_{10}$ and LaD$_{10}$ at 250, 300, and 350 GPa.}
\label{t1}
\begin{ruledtabular}
\begin{tabular}{c|c|c|c|c|c|c|c|c}
\multirow{2}{*}{$p$ (GPa)} &\multirow{2}{*}{$\mu^{\star}$} &\multirow{2}{*}{$\lambda$} &\multicolumn{2}{c|}{$T_c$ (K)} & \multicolumn{2}{c|}{$\Delta(0)$ (meV)~~} & \multicolumn{2}{c}{$2\Delta\left(0\right)/k_{B}T_{C}$}  \\ \cline{4-9}
                       &      &                        & ~~~LaH$_{10}$~~~~~  & ~~~LaD$_{10}$~~~~~     & ~~~LaH$_{10}$~~~~~     & ~~~LaD$_{10}$~~~~~    & ~~~LaH$_{10}$~~~~~     & ~~~LaD$_{10}$~~~~~     \\ \hline
                       & 0.10  &               & 269           & 191            & 59.74          & 42.80         &  5.15	        & 5.20           \\
250                    & 0.15  & 2.38          & 249           & 179            & 54.71          & 39.44         &  5.10	        & 5.11           \\
                       & 0.20  &               & 234           & 169            & 50.79          & 36.85         &  5.04	        & 5.06           \\ \hline
                       & 0.10  &               & 252           & 180            & 51.42          & 36.94         &  4.74	        & 4.76           \\
300                    & 0.15  & 1.82          & 231           & 166            & 46.30          & 33.52         &  4.65	        & 4.69           \\
                       & 0.20  &               & 214           & 155            & 42.33          & 30.90         &  4.59        & 4.63           \\ \hline
                       & 0.10  &               & 237           & 170            & 45.71          & 32.95         &  4.48	        & 4.50           \\
350                    & 0.15  & 1.54          & 214           & 154            & 40.55          & 29.49         &  4.40        & 4.44           \\
                       & 0.20  &               & 195           & 142            & 36.56          & 26.85         &  4.35        & 4.39           \\
\end{tabular}
\end{ruledtabular}
\end{table*}

\begin{figure}[!h]
\includegraphics[width=1\columnwidth]{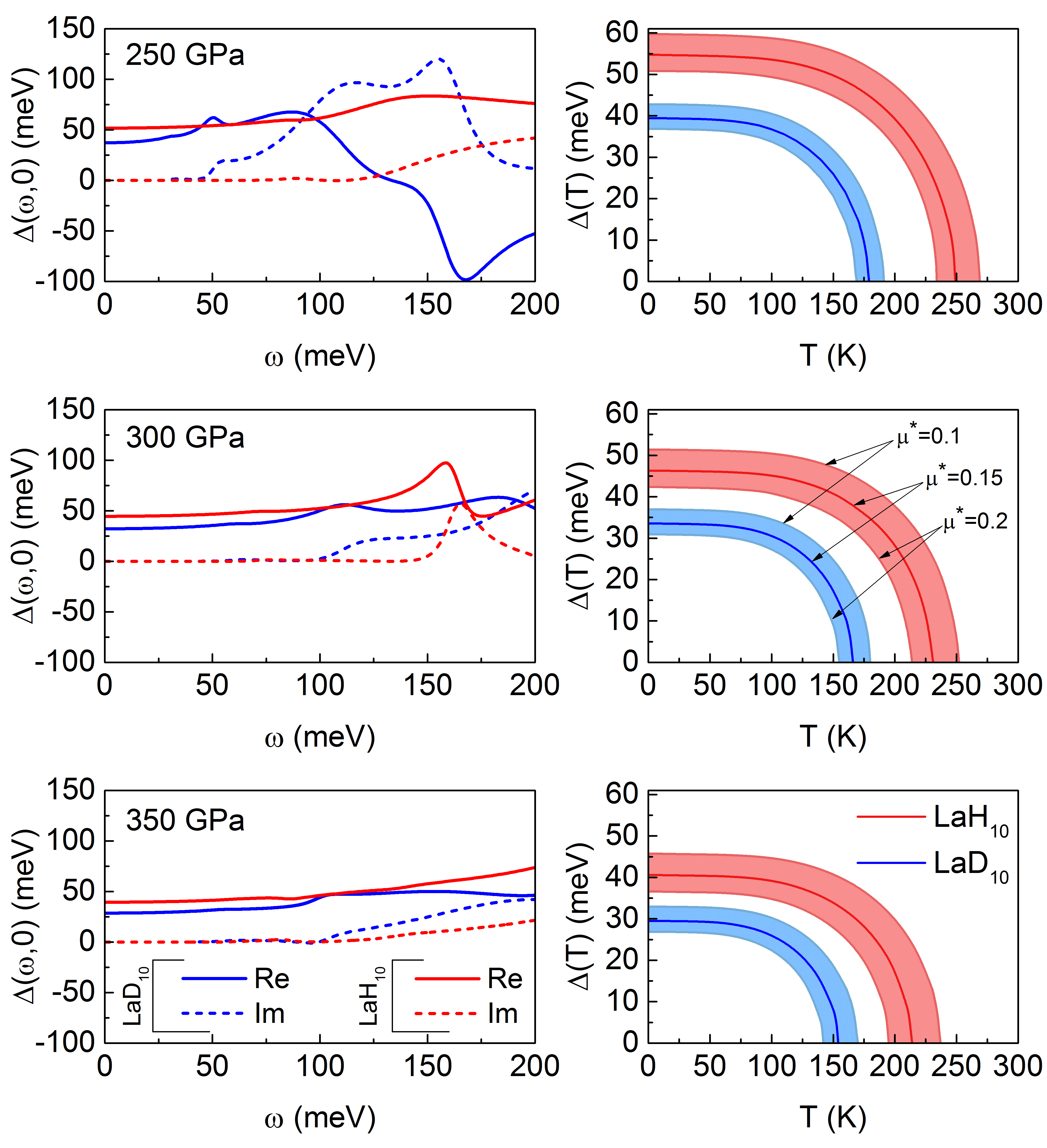}
\caption{Real and imaginary part of the superconducting energy gap as a function of frequency at zero temperature and $\mu^{\star}=0.15$ (left panel), and the corresponding superconducting energy gap as a function of temperature and Coulomb pseudopotential (right panel) determined from the relation $\Delta\left(T\right)={\rm Re}\left[\Delta\left(\omega=\Delta\left(T\right),T\right)\right]$.}
\label{f4}
\end{figure}

\begin{figure}[!h]
\includegraphics[width=0.9\columnwidth]{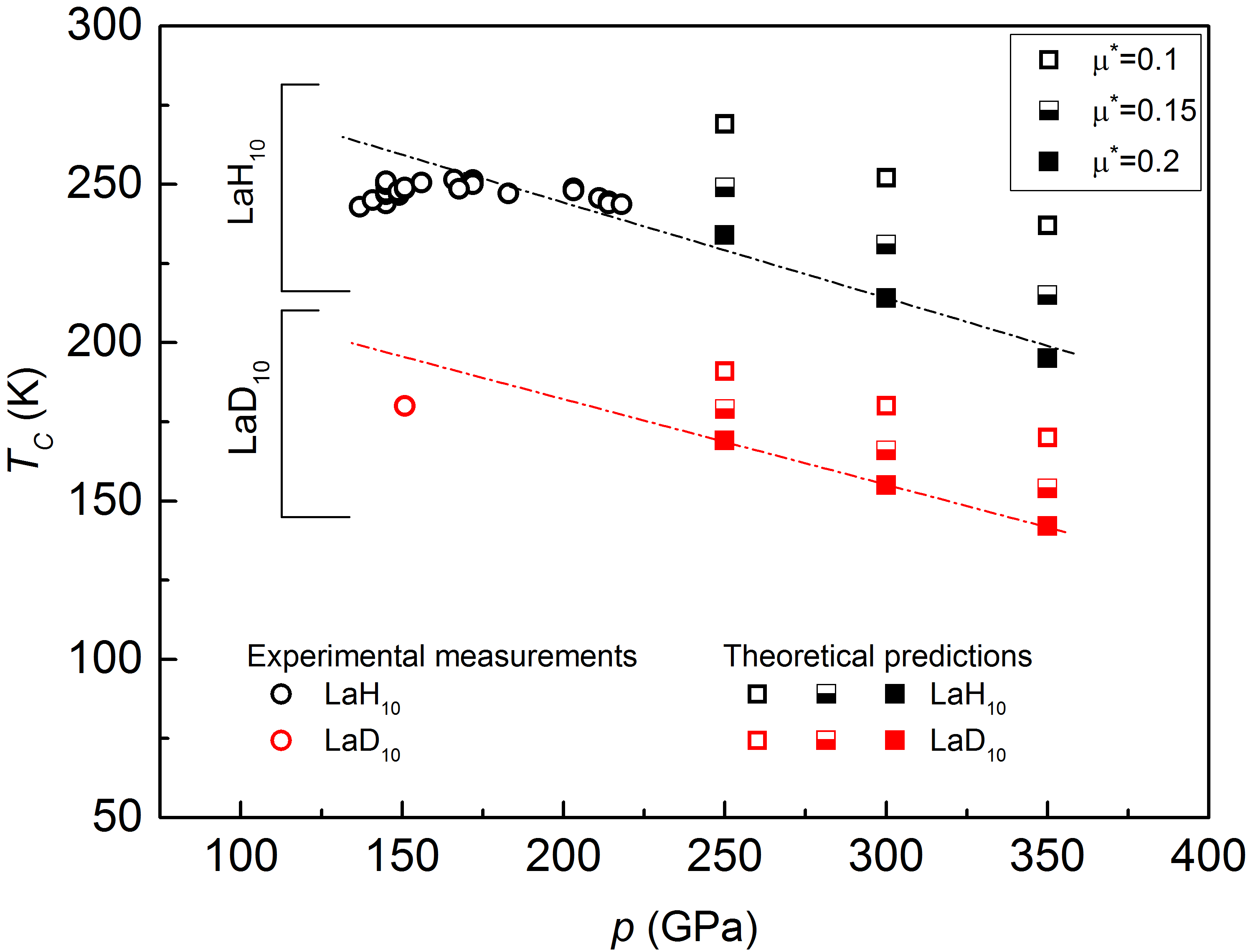}
\caption{Estimated $T_c$ as a function of pressure using the Eliashberg theory with $\mu^{\star}={0.1, 0.15, 0.2}$. The experimental data \cite{Drozdov2019} are also given for comparison.}
\label{f5}
\end{figure}

In Fig. 5, we compare the calculated $T_c$ $vs$ pressure results with the experimental data.
In the case of LaH$_{10}$, the open black circles represent the experimentally observed $T_c$ \cite{Drozdov2019}.
Due to the difficulty in the synthesis of fcc LaD$_{10}$, there was only one experimental data of $T_c$ measured at $150$ GPa (see a red open circle in Fig. 5) \cite{Drozdov2019}.
It is noteworthy that for LaH$_{10}$, the estimated $T_c$ decreases almost linearly with increasing pressure, irrespective of the chosen ${\mu}^{\star}$ values of $0.1$, $0.15$, and $0.2$.
This linear variation of $T_c$ with respect to pressure seems to be consistent with the experimental data.
Interestingly, we emphasize that H$_3$S and H$_3$S$_{1-x}$P$_x$ mixture \cite{Durajski-54-2017, Durajski-65-2018} also exhibited similar behavior of $T_c$ decrease with increasing pressure between $150$ and $350$ GPa.
As shown in Fig. 5, the calculated $T_c$ $vs$ pressure relation with ${\mu}^{\star}=0.2$ is in better agreement with the measurements (see the black dashed line).
For LaD$_{10}$, there was only one experimental data at $150$ GPa, which can be well extrapolated by our results obtained using ${\mu}^{\star}=0.2$ (see the red dashed line in Fig. 5).
It is noted that the experimental values of $T_c$ are well reproduced by the present harmonic approximation with ${\mu}^{\star}=0.2$, larger than that (${\mu}^{\star}=0.1$) of a recent anharmonic calculation \cite{Errea2020}. It is thus likely that the proper ${\mu}^{\star}$ value for LaH$_{10}$ would change with including anharmonic phonons.
The present theory and previous experimental data showing that $T_c$ varies with the isotopic mass provide a strong evidence for the interaction between the electrons and lattice vibrations \cite{AkashiH2S, Errea2015A, Mendez-Moreno}.
Indeed, this conventional electron-phonon coupling mechanism contrasts with the recent proposal that LaH$_{10}$ is an unconventional superconductor \cite{Talantsev_2019}.
We further analyze the isotope effect of $T_c$ in terms of the isotope coefficient ($\alpha$), which is given by the following relation:

\begin{equation}
\label{alpha}
\alpha=-\frac{ {\rm ln}[T_{C}]_{{\rm LaD}_{10}}-{\rm ln}[T_{C}]_{{\rm LaH}_{10}}}{{\rm ln}[M]_{\rm D}-{\rm ln}[M]_{\rm H}}.
\end{equation}
Here $[M]_{\rm H}$ and $[M]_{\rm D}$ are the atomic mass of hydrogen and deuterium, respectively.
It is noteworthy that the BCS theory for weak-coupling superconductors gives a $\alpha$ value of $0.5$, while the McMillan-Allen-Dynes or Migdal-Eliashberg equations with harmonic phonons have predicted $\alpha$ values lower than $0.5$ for strong-coupling superconductors: e.g., $0.23-0.31$ ($0.38-0.42$) for H$_2$S (H$_3$S) from McMillan-Allen-Dyne formula \cite{AkashiH2S} and $0.31$ for H$_3$S from Migdal-Eliashberg equations \cite{SZCZESNIAK201730}.

As shown in \tab{t2}, the $\alpha$ value ranges between $0.458$ and $0.470$, which are smaller than the BCS value which indicate the correction due to the strong coupling and retardation effects.
Moreover, on the basis of only one available experimental data of $T_c$ for LaH$_{10}$ ($249$ K) and LaD$_{10}$ ($180$ K) at a pressure of ${\sim}150$ GPa, the ${\alpha}$ value can be estimated as $0.46$, in good agreement with our estimated average isotope effect coefficient $\overline{\alpha}=0.465$.
We note that the present value of $\alpha$ estimated at higher pressures between $250$ and $350$ GPa agrees well with the previous theoretical value ($0.43$ at $\sim$160 GPa) obtained using anharmonic phonons \cite{Errea2020}. It is thus likely that LaH$_{10}$ exhibits insignificant anharmonic effects for $\alpha$.

It is noteworthy that a recent anisotropic Migdal-Eliashberg theory with including anharmonic effect predicted $T_c$ at pressures lower than $260$ GPa in the Fm$\overline{3}$m structure \cite{Errea2020}. As shown in Fig. S2(a) of the Supplemental Material \cite{SM}, the predicted $T_c$ vs pressure results \cite{Errea2020} agree well with the experimental data \cite{Drozdov2019} measured between $150$ and $210$ GPa. Here, the theoretical values of $T_c$ decrease by $\sim$21 K between $214$ and $264$ GPa, close to a slope of $\sim$15 K per $50$ GPa between $250$ and $350$ GPa (see Fig. 5 or Fig. S2(b) in the Supplemental Material \cite{SM}) which is obtained using isotropic Migdal-Eliashberg equations with $\mu^{\star}=0.2$ and harmonic phonons. It is thus likely that the pressure dependence of $T_c$ at high pressures above $220$ GPa shows good agreement between the previous anharmonic \cite{Errea2020} and the present harmonic calculations. We note that the present isotope coefficient of $0.465$ estimated at higher pressures between $250$ and $350$ GPa agree well with the previous theoretical value ($0.43$ at $\sim$160 GPa) obtained using anharmonic phonons \cite{Errea2020} as well as the existing experimental data ($\alpha = 0.46$) measured only at a pressure of $\sim$150 GPa \cite{Drozdov2019}. This agreement of $\alpha$ between the previous anharmonic \cite{Errea2020} and the present harmonic calculations is due to the fact that the two calculations predict similar $T_c$ $vs$ pressure results for LaH$_{10}$ and LaD$_{10}$, as shown in Fig. S2(a) and S2(b) of the Supplemental Material \cite{SM}. Our findings not only support a conventional electron-phonon coupling mechanism in the superconductivity of fcc LaH$_{10}$, but also will stimulate further experiments to explore the isotope effect of $T_c$ at higher pressures.

\begin{table}[!h]
\centering
\caption{Estimated $T_c$ with $\mu^{\star}=0.2$, the isotope effect coefficient $\alpha$, and the average isotope effect coefficient $\overline{\alpha}$ of fcc LaH$_{10}$ and LaD$_{10}$ at 250, 300, and 350 GPa.}
\label{t2}
\begin{ruledtabular}
\begin{tabular}{c|c|c|c|c}
\multirow{2}{*}{pressure (GPa)} & \multicolumn{2}{c|}{$T_c$ (K)} & \multirow{2}{*}{$\alpha$~~} & \multirow{2}{*}{$\overline{\alpha}$}  \\ \cline{2-3}
                         & ~~~~LaH$_{10}$~~~~     & LaD$_{10}$    &                &        \\ \hline
250                      & 234           & 169            & 0.470          &        \\
300                      & 214           & 155            & 0.466          & 0.465  \\
350                      & 195           & 142            & 0.458          &        \\
\end{tabular}
\end{ruledtabular}
\end{table}

\section{IV. Conclusions}

We have performed the comprehensive first-principles DFT calculations of the electronic and phonon properties of lanthanum hydride and deuteride at high pressures.
We found that fcc LaH$_{10}$ and LaD$_{10}$ phases not only dynamically stabilize in the range of $250-350$ GPa, but also have the strong EPC and the high DOS at the Fermi level which are favorable for high-$T_c$ SC.
By solving the Eliashberg equations, $T_c$ of LaH$_{10}$ (LaD$_{10}$) decreases almost linearly as 234 (169), 214 (155), 195 (142) K at 250, 300, and 350 GPa, respectively. As a result, the isotope coefficient is estimated to be $\sim$0.465, irrespective of pressure. Therefore, our findings strongly support a conventional electron-phonon coupling mechanism in the observed room-temperature superconductivity of fcc LaH$_{10}$.
We believe that these results are of importance for understanding the phenomenon of superconducting state in fcc LaH$_{10}$, which will be useful for future more experimental measurements of isotope effect.

\section{Acknowledgements}

Artur P. Durajski acknowledges the financial support from the Polish National Science Centre (NCN) under grant No. 2016/23/D/ST3/02109 and from the Polish Ministry of Science and Higher Education under the scholarship for young outstanding scientists No. 406/STYP/13/2018. Jun-Hyung Cho acknowledges the National Research Foundation of Korea (NRF) grant funded by the Korean Government (Grants No. 2019R1A2C1002975, No. 2016K1A4A3914691, and No. 2015M3D1A1070609).
Yinwei Li acknowledges funding from the National Natural Science Foundation of China under Grant No. 11722433, the Six Talent Peaks Project of Jiangsu Province.
%

\bibliographystyle{apsrev4-1}
\bibliography{bibliography}

\begin{thebibliography}{55}%
\makeatletter
\providecommand \@ifxundefined [1]{%
 \@ifx{#1\undefined}
}%
\providecommand \@ifnum [1]{%
 \ifnum #1\expandafter \@firstoftwo
 \else \expandafter \@secondoftwo
 \fi
}%
\providecommand \@ifx [1]{%
 \ifx #1\expandafter \@firstoftwo
 \else \expandafter \@secondoftwo
 \fi
}%
\providecommand \natexlab [1]{#1}%
\providecommand \enquote  [1]{``#1''}%
\providecommand \bibnamefont  [1]{#1}%
\providecommand \bibfnamefont [1]{#1}%
\providecommand \citenamefont [1]{#1}%
\providecommand \href@noop [0]{\@secondoftwo}%
\providecommand \href [0]{\begingroup \@sanitize@url \@href}%
\providecommand \@href[1]{\@@startlink{#1}\@@href}%
\providecommand \@@href[1]{\endgroup#1\@@endlink}%
\providecommand \@sanitize@url [0]{\catcode `\\12\catcode `\$12\catcode
  `\&12\catcode `\#12\catcode `\^12\catcode `\_12\catcode `\%12\relax}%
\providecommand \@@startlink[1]{}%
\providecommand \@@endlink[0]{}%
\providecommand \url  [0]{\begingroup\@sanitize@url \@url }%
\providecommand \@url [1]{\endgroup\@href {#1}{\urlprefix }}%
\providecommand \urlprefix  [0]{URL }%
\providecommand \Eprint [0]{\href }%
\providecommand \doibase [0]{http://dx.doi.org/}%
\providecommand \selectlanguage [0]{\@gobble}%
\providecommand \bibinfo  [0]{\@secondoftwo}%
\providecommand \bibfield  [0]{\@secondoftwo}%
\providecommand \translation [1]{[#1]}%
\providecommand \BibitemOpen [0]{}%
\providecommand \bibitemStop [0]{}%
\providecommand \bibitemNoStop [0]{.\EOS\space}%
\providecommand \EOS [0]{\spacefactor3000\relax}%
\providecommand \BibitemShut  [1]{\csname bibitem#1\endcsname}%
\let\auto@bib@innerbib\@empty
\bibitem [{\citenamefont {Onnes}(1911)}]{onnes}%
  \BibitemOpen
  \bibfield  {author} {\bibinfo {author} {\bibfnamefont {H.~K.}\ \bibnamefont
  {Onnes}},\ }\href@noop {} {\bibfield  {journal} {\bibinfo  {journal} {Comm.
  Phys. Lab. Univ. Leiden}\ }\textbf {\bibinfo {volume} {122}},\ \bibinfo
  {pages} {124} (\bibinfo {year} {1911})}\BibitemShut {NoStop}%
\bibitem [{\citenamefont {Bardeen}\ \emph {et~al.}(1957)\citenamefont
  {Bardeen}, \citenamefont {Cooper},\ and\ \citenamefont
  {Schrieffer}}]{Bardeen1957A}%
  \BibitemOpen
  \bibfield  {author} {\bibinfo {author} {\bibfnamefont {J.}~\bibnamefont
  {Bardeen}}, \bibinfo {author} {\bibfnamefont {L.~N.}\ \bibnamefont {Cooper}},
  \ and\ \bibinfo {author} {\bibfnamefont {J.~R.}\ \bibnamefont {Schrieffer}},\
  }\href {\doibase 10.1103/PhysRev.106.162} {\bibfield  {journal} {\bibinfo
  {journal} {Phys. Rev.}\ }\textbf {\bibinfo {volume} {106}},\ \bibinfo {pages}
  {162} (\bibinfo {year} {1957})}\BibitemShut {NoStop}%
\bibitem [{\citenamefont {Ashcroft}(1968)}]{Ashcroft1968}%
  \BibitemOpen
  \bibfield  {author} {\bibinfo {author} {\bibfnamefont {N.~W.}\ \bibnamefont
  {Ashcroft}},\ }\href {\doibase 10.1103/PhysRevLett.21.1748} {\bibfield
  {journal} {\bibinfo  {journal} {Phys. Rev. Lett.}\ }\textbf {\bibinfo
  {volume} {21}},\ \bibinfo {pages} {1748} (\bibinfo {year}
  {1968})}\BibitemShut {NoStop}%
\bibitem [{\citenamefont {McMinis}\ \emph {et~al.}(2015)\citenamefont
  {McMinis}, \citenamefont {Clay}, \citenamefont {Lee},\ and\ \citenamefont
  {Morales}}]{MetalicH1}%
  \BibitemOpen
  \bibfield  {author} {\bibinfo {author} {\bibfnamefont {J.}~\bibnamefont
  {McMinis}}, \bibinfo {author} {\bibfnamefont {R.~C.}\ \bibnamefont {Clay}},
  \bibinfo {author} {\bibfnamefont {D.}~\bibnamefont {Lee}}, \ and\ \bibinfo
  {author} {\bibfnamefont {M.~A.}\ \bibnamefont {Morales}},\ }\href {\doibase
  10.1103/PhysRevLett.114.105305} {\bibfield  {journal} {\bibinfo  {journal}
  {Phys. Rev. Lett.}\ }\textbf {\bibinfo {volume} {114}},\ \bibinfo {pages}
  {105305} (\bibinfo {year} {2015})}\BibitemShut {NoStop}%
\bibitem [{\citenamefont {McMahon}\ \emph {et~al.}(2012)\citenamefont
  {McMahon}, \citenamefont {Morales}, \citenamefont {Pierleoni},\ and\
  \citenamefont {Ceperley}}]{MetalicH2}%
  \BibitemOpen
  \bibfield  {author} {\bibinfo {author} {\bibfnamefont {J.~M.}\ \bibnamefont
  {McMahon}}, \bibinfo {author} {\bibfnamefont {M.~A.}\ \bibnamefont
  {Morales}}, \bibinfo {author} {\bibfnamefont {C.}~\bibnamefont {Pierleoni}},
  \ and\ \bibinfo {author} {\bibfnamefont {D.~M.}\ \bibnamefont {Ceperley}},\
  }\href {\doibase 10.1103/RevModPhys.84.1607} {\bibfield  {journal} {\bibinfo
  {journal} {Rev. Mod. Phys.}\ }\textbf {\bibinfo {volume} {84}},\ \bibinfo
  {pages} {1607} (\bibinfo {year} {2012})}\BibitemShut {NoStop}%
\bibitem [{\citenamefont {Bassett}(2009)}]{diamondanvil1}%
  \BibitemOpen
  \bibfield  {author} {\bibinfo {author} {\bibfnamefont {W.~A.}\ \bibnamefont
  {Bassett}},\ }\href {\doibase 10.1080/08957950802597239} {\bibfield
  {journal} {\bibinfo  {journal} {High Press. Res.}\ }\textbf {\bibinfo
  {volume} {29}},\ \bibinfo {pages} {163} (\bibinfo {year} {2009})}\BibitemShut
  {NoStop}%
\bibitem [{\citenamefont {Mao}\ \emph {et~al.}(2018)\citenamefont {Mao},
  \citenamefont {Chen}, \citenamefont {Ding}, \citenamefont {Li},\ and\
  \citenamefont {Wang}}]{diamondanvil2}%
  \BibitemOpen
  \bibfield  {author} {\bibinfo {author} {\bibfnamefont {H.-K.}\ \bibnamefont
  {Mao}}, \bibinfo {author} {\bibfnamefont {X.-J.}\ \bibnamefont {Chen}},
  \bibinfo {author} {\bibfnamefont {Y.}~\bibnamefont {Ding}}, \bibinfo {author}
  {\bibfnamefont {B.}~\bibnamefont {Li}}, \ and\ \bibinfo {author}
  {\bibfnamefont {L.}~\bibnamefont {Wang}},\ }\href {\doibase
  10.1103/RevModPhys.90.015007} {\bibfield  {journal} {\bibinfo  {journal}
  {Rev. Mod. Phys.}\ }\textbf {\bibinfo {volume} {90}},\ \bibinfo {pages}
  {015007} (\bibinfo {year} {2018})}\BibitemShut {NoStop}%
\bibitem [{\citenamefont {Wang}\ \emph {et~al.}(2012)\citenamefont {Wang},
  \citenamefont {Tse}, \citenamefont {Tanaka}, \citenamefont {Iitaka},\ and\
  \citenamefont {Ma}}]{Wang2012-CaH6}%
  \BibitemOpen
  \bibfield  {author} {\bibinfo {author} {\bibfnamefont {H.}~\bibnamefont
  {Wang}}, \bibinfo {author} {\bibfnamefont {J.~S.}\ \bibnamefont {Tse}},
  \bibinfo {author} {\bibfnamefont {K.}~\bibnamefont {Tanaka}}, \bibinfo
  {author} {\bibfnamefont {T.}~\bibnamefont {Iitaka}}, \ and\ \bibinfo {author}
  {\bibfnamefont {Y.}~\bibnamefont {Ma}},\ }\href {\doibase
  10.1073/pnas.1118168109} {\bibfield  {journal} {\bibinfo  {journal} {Proc.
  Natl. Acad. Sci. USA}\ }\textbf {\bibinfo {volume} {109}},\ \bibinfo {pages}
  {6463} (\bibinfo {year} {2012})}\BibitemShut {NoStop}%
\bibitem [{\citenamefont {Duan}\ \emph {et~al.}(2014)\citenamefont {Duan},
  \citenamefont {Liu}, \citenamefont {Tian}, \citenamefont {Li}, \citenamefont
  {Huang}, \citenamefont {Zhao}, \citenamefont {Yu}, \citenamefont {Liu},
  \citenamefont {Tian},\ and\ \citenamefont {Cui}}]{Duan2014A}%
  \BibitemOpen
  \bibfield  {author} {\bibinfo {author} {\bibfnamefont {D.}~\bibnamefont
  {Duan}}, \bibinfo {author} {\bibfnamefont {Y.}~\bibnamefont {Liu}}, \bibinfo
  {author} {\bibfnamefont {F.}~\bibnamefont {Tian}}, \bibinfo {author}
  {\bibfnamefont {D.}~\bibnamefont {Li}}, \bibinfo {author} {\bibfnamefont
  {X.}~\bibnamefont {Huang}}, \bibinfo {author} {\bibfnamefont
  {Z.}~\bibnamefont {Zhao}}, \bibinfo {author} {\bibfnamefont {H.}~\bibnamefont
  {Yu}}, \bibinfo {author} {\bibfnamefont {B.}~\bibnamefont {Liu}}, \bibinfo
  {author} {\bibfnamefont {W.}~\bibnamefont {Tian}}, \ and\ \bibinfo {author}
  {\bibfnamefont {T.}~\bibnamefont {Cui}},\ }\href
  {http://dx.doi.org/10.1038/srep06968} {\bibfield  {journal} {\bibinfo
  {journal} {Sci. Rep.}\ }\textbf {\bibinfo {volume} {4}},\ \bibinfo {pages}
  {6968} (\bibinfo {year} {2014})}\BibitemShut {NoStop}%
\bibitem [{\citenamefont {Feng}\ \emph {et~al.}(2015)\citenamefont {Feng},
  \citenamefont {Zhang}, \citenamefont {Gao}, \citenamefont {Liu},\ and\
  \citenamefont {Wang}}]{Feng2015-MgH6}%
  \BibitemOpen
  \bibfield  {author} {\bibinfo {author} {\bibfnamefont {X.}~\bibnamefont
  {Feng}}, \bibinfo {author} {\bibfnamefont {J.}~\bibnamefont {Zhang}},
  \bibinfo {author} {\bibfnamefont {G.}~\bibnamefont {Gao}}, \bibinfo {author}
  {\bibfnamefont {H.}~\bibnamefont {Liu}}, \ and\ \bibinfo {author}
  {\bibfnamefont {H.}~\bibnamefont {Wang}},\ }\href {\doibase
  10.1039/C5RA11459D} {\bibfield  {journal} {\bibinfo  {journal} {RSC Adv.}\
  }\textbf {\bibinfo {volume} {5}},\ \bibinfo {pages} {59292} (\bibinfo {year}
  {2015})}\BibitemShut {NoStop}%
\bibitem [{\citenamefont {Peng}\ \emph {et~al.}(2017)\citenamefont {Peng},
  \citenamefont {Sun}, \citenamefont {Pickard}, \citenamefont {Needs},
  \citenamefont {Wu},\ and\ \citenamefont {Ma}}]{rare-earth-hydride1}%
  \BibitemOpen
  \bibfield  {author} {\bibinfo {author} {\bibfnamefont {F.}~\bibnamefont
  {Peng}}, \bibinfo {author} {\bibfnamefont {Y.}~\bibnamefont {Sun}}, \bibinfo
  {author} {\bibfnamefont {C.~J.}\ \bibnamefont {Pickard}}, \bibinfo {author}
  {\bibfnamefont {R.~J.}\ \bibnamefont {Needs}}, \bibinfo {author}
  {\bibfnamefont {Q.}~\bibnamefont {Wu}}, \ and\ \bibinfo {author}
  {\bibfnamefont {Y.}~\bibnamefont {Ma}},\ }\href {\doibase
  10.1103/PhysRevLett.119.107001} {\bibfield  {journal} {\bibinfo  {journal}
  {Phys. Rev. Lett.}\ }\textbf {\bibinfo {volume} {119}},\ \bibinfo {pages}
  {107001} (\bibinfo {year} {2017})}\BibitemShut {NoStop}%
\bibitem [{\citenamefont {Liu}\ \emph {et~al.}(2017)\citenamefont {Liu},
  \citenamefont {Naumov}, \citenamefont {Hoffmann}, \citenamefont {Ashcroft},\
  and\ \citenamefont {Hemley}}]{rare-earth-hydride2}%
  \BibitemOpen
  \bibfield  {author} {\bibinfo {author} {\bibfnamefont {H.}~\bibnamefont
  {Liu}}, \bibinfo {author} {\bibfnamefont {I.~I.}\ \bibnamefont {Naumov}},
  \bibinfo {author} {\bibfnamefont {R.}~\bibnamefont {Hoffmann}}, \bibinfo
  {author} {\bibfnamefont {N.~W.}\ \bibnamefont {Ashcroft}}, \ and\ \bibinfo
  {author} {\bibfnamefont {R.~J.}\ \bibnamefont {Hemley}},\ }\href {\doibase
  10.1073/pnas.1704505114} {\bibfield  {journal} {\bibinfo  {journal} {Proc.
  Natl. Acad. Sci. USA}\ }\textbf {\bibinfo {volume} {114}},\ \bibinfo {pages}
  {6990} (\bibinfo {year} {2017})}\BibitemShut {NoStop}%
\bibitem [{\citenamefont {Zurek}\ and\ \citenamefont {Bi}(2019)}]{Hydride1}%
  \BibitemOpen
  \bibfield  {author} {\bibinfo {author} {\bibfnamefont {E.}~\bibnamefont
  {Zurek}}\ and\ \bibinfo {author} {\bibfnamefont {T.}~\bibnamefont {Bi}},\
  }\href {\doibase 10.1063/1.5079225} {\bibfield  {journal} {\bibinfo
  {journal} {J. Chem. Phys.}\ }\textbf {\bibinfo {volume} {150}},\ \bibinfo
  {pages} {050901} (\bibinfo {year} {2019})}\BibitemShut {NoStop}%
\bibitem [{\citenamefont {Flores-Livas}\ \emph {et~al.}(2020)\citenamefont
  {Flores-Livas}, \citenamefont {Boeri}, \citenamefont {Sanna}, \citenamefont
  {Profeta}, \citenamefont {Arita},\ and\ \citenamefont {Eremets}}]{Hydride2}%
  \BibitemOpen
  \bibfield  {author} {\bibinfo {author} {\bibfnamefont {J.~A.}\ \bibnamefont
  {Flores-Livas}}, \bibinfo {author} {\bibfnamefont {L.}~\bibnamefont {Boeri}},
  \bibinfo {author} {\bibfnamefont {A.}~\bibnamefont {Sanna}}, \bibinfo
  {author} {\bibfnamefont {G.}~\bibnamefont {Profeta}}, \bibinfo {author}
  {\bibfnamefont {R.}~\bibnamefont {Arita}}, \ and\ \bibinfo {author}
  {\bibfnamefont {M.}~\bibnamefont {Eremets}},\ }\href {\doibase
  https://doi.org/10.1016/j.physrep.2020.02.003} {\bibfield  {journal}
  {\bibinfo  {journal} {Phys. Rep.}\ }\textbf {\bibinfo {volume} {856}},\
  \bibinfo {pages} {1 } (\bibinfo {year} {2020})}\BibitemShut {NoStop}%
\bibitem [{\citenamefont {Li}\ \emph {et~al.}(2015)\citenamefont {Li},
  \citenamefont {Hao}, \citenamefont {Liu}, \citenamefont {Tse}, \citenamefont
  {Wang},\ and\ \citenamefont {Ma}}]{LiY2015}%
  \BibitemOpen
  \bibfield  {author} {\bibinfo {author} {\bibfnamefont {Y.}~\bibnamefont
  {Li}}, \bibinfo {author} {\bibfnamefont {J.}~\bibnamefont {Hao}}, \bibinfo
  {author} {\bibfnamefont {H.}~\bibnamefont {Liu}}, \bibinfo {author}
  {\bibfnamefont {J.~S.}\ \bibnamefont {Tse}}, \bibinfo {author} {\bibfnamefont
  {Y.}~\bibnamefont {Wang}}, \ and\ \bibinfo {author} {\bibfnamefont
  {Y.}~\bibnamefont {Ma}},\ }\href {\doibase 10.1038/srep09948} {\bibfield
  {journal} {\bibinfo  {journal} {Sci. Rep.}\ }\textbf {\bibinfo {volume}
  {5}},\ \bibinfo {pages} {9948} (\bibinfo {year} {2015})}\BibitemShut
  {NoStop}%
\bibitem [{\citenamefont {Li}\ \emph {et~al.}(2016)\citenamefont {Li},
  \citenamefont {Wang}, \citenamefont {Liu}, \citenamefont {Zhang},
  \citenamefont {Hao}, \citenamefont {Pickard}, \citenamefont {Nelson},
  \citenamefont {Needs}, \citenamefont {Li}, \citenamefont {Huang},
  \citenamefont {Errea}, \citenamefont {Calandra}, \citenamefont {Mauri},\ and\
  \citenamefont {Ma}}]{PhysRevB.93.020103}%
  \BibitemOpen
  \bibfield  {author} {\bibinfo {author} {\bibfnamefont {Y.}~\bibnamefont
  {Li}}, \bibinfo {author} {\bibfnamefont {L.}~\bibnamefont {Wang}}, \bibinfo
  {author} {\bibfnamefont {H.}~\bibnamefont {Liu}}, \bibinfo {author}
  {\bibfnamefont {Y.}~\bibnamefont {Zhang}}, \bibinfo {author} {\bibfnamefont
  {J.}~\bibnamefont {Hao}}, \bibinfo {author} {\bibfnamefont {C.~J.}\
  \bibnamefont {Pickard}}, \bibinfo {author} {\bibfnamefont {J.~R.}\
  \bibnamefont {Nelson}}, \bibinfo {author} {\bibfnamefont {R.~J.}\
  \bibnamefont {Needs}}, \bibinfo {author} {\bibfnamefont {W.}~\bibnamefont
  {Li}}, \bibinfo {author} {\bibfnamefont {Y.}~\bibnamefont {Huang}}, \bibinfo
  {author} {\bibfnamefont {I.}~\bibnamefont {Errea}}, \bibinfo {author}
  {\bibfnamefont {M.}~\bibnamefont {Calandra}}, \bibinfo {author}
  {\bibfnamefont {F.}~\bibnamefont {Mauri}}, \ and\ \bibinfo {author}
  {\bibfnamefont {Y.}~\bibnamefont {Ma}},\ }\href {\doibase
  10.1103/PhysRevB.93.020103} {\bibfield  {journal} {\bibinfo  {journal} {Phys.
  Rev. B}\ }\textbf {\bibinfo {volume} {93}},\ \bibinfo {pages} {020103(R)}
  (\bibinfo {year} {2016})}\BibitemShut {NoStop}%
\bibitem [{\citenamefont {Quan}\ \emph {et~al.}(2019)\citenamefont {Quan},
  \citenamefont {Ghosh},\ and\ \citenamefont {Pickett}}]{QuanYundi}%
  \BibitemOpen
  \bibfield  {author} {\bibinfo {author} {\bibfnamefont {Y.}~\bibnamefont
  {Quan}}, \bibinfo {author} {\bibfnamefont {S.~S.}\ \bibnamefont {Ghosh}}, \
  and\ \bibinfo {author} {\bibfnamefont {W.~E.}\ \bibnamefont {Pickett}},\
  }\href {\doibase 10.1103/PhysRevB.100.184505} {\bibfield  {journal} {\bibinfo
   {journal} {Phys. Rev. B}\ }\textbf {\bibinfo {volume} {100}},\ \bibinfo
  {pages} {184505} (\bibinfo {year} {2019})}\BibitemShut {NoStop}%
\bibitem [{\citenamefont {Liu}\ \emph {et~al.}(2018)\citenamefont {Liu},
  \citenamefont {Naumov}, \citenamefont {Geballe}, \citenamefont {Somayazulu},
  \citenamefont {Tse},\ and\ \citenamefont {Hemley}}]{PhysRevB.98.100102}%
  \BibitemOpen
  \bibfield  {author} {\bibinfo {author} {\bibfnamefont {H.}~\bibnamefont
  {Liu}}, \bibinfo {author} {\bibfnamefont {I.~I.}\ \bibnamefont {Naumov}},
  \bibinfo {author} {\bibfnamefont {Z.~M.}\ \bibnamefont {Geballe}}, \bibinfo
  {author} {\bibfnamefont {M.}~\bibnamefont {Somayazulu}}, \bibinfo {author}
  {\bibfnamefont {J.~S.}\ \bibnamefont {Tse}}, \ and\ \bibinfo {author}
  {\bibfnamefont {R.~J.}\ \bibnamefont {Hemley}},\ }\href {\doibase
  10.1103/PhysRevB.98.100102} {\bibfield  {journal} {\bibinfo  {journal} {Phys.
  Rev. B}\ }\textbf {\bibinfo {volume} {98}},\ \bibinfo {pages} {100102(R)}
  (\bibinfo {year} {2018})}\BibitemShut {NoStop}%
\bibitem [{\citenamefont {Li}\ \emph {et~al.}(2014)\citenamefont {Li},
  \citenamefont {Hao}, \citenamefont {Liu}, \citenamefont {Li},\ and\
  \citenamefont {Ma}}]{Li2014A}%
  \BibitemOpen
  \bibfield  {author} {\bibinfo {author} {\bibfnamefont {Y.}~\bibnamefont
  {Li}}, \bibinfo {author} {\bibfnamefont {J.}~\bibnamefont {Hao}}, \bibinfo
  {author} {\bibfnamefont {H.}~\bibnamefont {Liu}}, \bibinfo {author}
  {\bibfnamefont {Y.}~\bibnamefont {Li}}, \ and\ \bibinfo {author}
  {\bibfnamefont {Y.}~\bibnamefont {Ma}},\ }\href {\doibase
  http://dx.doi.org/10.1063/1.4874158} {\bibfield  {journal} {\bibinfo
  {journal} {J. Chem. Phys.}\ }\textbf {\bibinfo {volume} {140}},\ \bibinfo
  {pages} {174712} (\bibinfo {year} {2014})}\BibitemShut {NoStop}%
\bibitem [{\citenamefont {Drozdov}\ \emph {et~al.}(2015)\citenamefont
  {Drozdov}, \citenamefont {Eremets}, \citenamefont {Troyan}, \citenamefont
  {Ksenofontov},\ and\ \citenamefont {Shylin}}]{Drozdov2015A}%
  \BibitemOpen
  \bibfield  {author} {\bibinfo {author} {\bibfnamefont {A.~P.}\ \bibnamefont
  {Drozdov}}, \bibinfo {author} {\bibfnamefont {M.~I.}\ \bibnamefont
  {Eremets}}, \bibinfo {author} {\bibfnamefont {I.~A.}\ \bibnamefont {Troyan}},
  \bibinfo {author} {\bibfnamefont {V.}~\bibnamefont {Ksenofontov}}, \ and\
  \bibinfo {author} {\bibfnamefont {S.~I.}\ \bibnamefont {Shylin}},\ }\href
  {\doibase 10.1038/nature14964} {\bibfield  {journal} {\bibinfo  {journal}
  {Nature}\ }\textbf {\bibinfo {volume} {525}},\ \bibinfo {pages} {73}
  (\bibinfo {year} {2015})}\BibitemShut {NoStop}%
\bibitem [{\citenamefont {Einaga}\ \emph {et~al.}(2016)\citenamefont {Einaga},
  \citenamefont {Sakata}, \citenamefont {Ishikawa}, \citenamefont {Shimizu},
  \citenamefont {Eremets}, \citenamefont {Drozdov}, \citenamefont {Troyan},
  \citenamefont {Hirao},\ and\ \citenamefont {Ohishi}}]{Einaga2016A}%
  \BibitemOpen
  \bibfield  {author} {\bibinfo {author} {\bibfnamefont {M.}~\bibnamefont
  {Einaga}}, \bibinfo {author} {\bibfnamefont {M.}~\bibnamefont {Sakata}},
  \bibinfo {author} {\bibfnamefont {T.}~\bibnamefont {Ishikawa}}, \bibinfo
  {author} {\bibfnamefont {K.}~\bibnamefont {Shimizu}}, \bibinfo {author}
  {\bibfnamefont {M.~I.}\ \bibnamefont {Eremets}}, \bibinfo {author}
  {\bibfnamefont {A.~P.}\ \bibnamefont {Drozdov}}, \bibinfo {author}
  {\bibfnamefont {I.~A.}\ \bibnamefont {Troyan}}, \bibinfo {author}
  {\bibfnamefont {N.}~\bibnamefont {Hirao}}, \ and\ \bibinfo {author}
  {\bibfnamefont {Y.}~\bibnamefont {Ohishi}},\ }\href
  {http://dx.doi.org/10.1038/nphys3760} {\bibfield  {journal} {\bibinfo
  {journal} {Nat. Phys.}\ }\textbf {\bibinfo {volume} {12}},\ \bibinfo {pages}
  {835} (\bibinfo {year} {2016})}\BibitemShut {NoStop}%
\bibitem [{\citenamefont {Somayazulu}\ \emph {et~al.}(2019)\citenamefont
  {Somayazulu}, \citenamefont {Ahart}, \citenamefont {Mishra}, \citenamefont
  {Geballe}, \citenamefont {Baldini}, \citenamefont {Meng}, \citenamefont
  {Struzhkin},\ and\ \citenamefont {Hemley}}]{ExpLaH10-PRL}%
  \BibitemOpen
  \bibfield  {author} {\bibinfo {author} {\bibfnamefont {M.}~\bibnamefont
  {Somayazulu}}, \bibinfo {author} {\bibfnamefont {M.}~\bibnamefont {Ahart}},
  \bibinfo {author} {\bibfnamefont {A.~K.}\ \bibnamefont {Mishra}}, \bibinfo
  {author} {\bibfnamefont {Z.~M.}\ \bibnamefont {Geballe}}, \bibinfo {author}
  {\bibfnamefont {M.}~\bibnamefont {Baldini}}, \bibinfo {author} {\bibfnamefont
  {Y.}~\bibnamefont {Meng}}, \bibinfo {author} {\bibfnamefont {V.~V.}\
  \bibnamefont {Struzhkin}}, \ and\ \bibinfo {author} {\bibfnamefont {R.~J.}\
  \bibnamefont {Hemley}},\ }\href {\doibase 10.1103/PhysRevLett.122.027001}
  {\bibfield  {journal} {\bibinfo  {journal} {Phys. Rev. Lett.}\ }\textbf
  {\bibinfo {volume} {122}},\ \bibinfo {pages} {027001} (\bibinfo {year}
  {2019})}\BibitemShut {NoStop}%
\bibitem [{\citenamefont {Drozdov}\ \emph {et~al.}(2019)\citenamefont
  {Drozdov}, \citenamefont {Kong}, \citenamefont {Minkov}, \citenamefont
  {Besedin}, \citenamefont {Kuzovnikov}, \citenamefont {Mozaffari},
  \citenamefont {Balicas}, \citenamefont {Balakirev}, \citenamefont {Graf},
  \citenamefont {Prakapenka}, \citenamefont {Greenberg}, \citenamefont
  {Knyazev}, \citenamefont {Tkacz},\ and\ \citenamefont
  {Eremets}}]{Drozdov2019}%
  \BibitemOpen
  \bibfield  {author} {\bibinfo {author} {\bibfnamefont {A.~P.}\ \bibnamefont
  {Drozdov}}, \bibinfo {author} {\bibfnamefont {P.~P.}\ \bibnamefont {Kong}},
  \bibinfo {author} {\bibfnamefont {V.~S.}\ \bibnamefont {Minkov}}, \bibinfo
  {author} {\bibfnamefont {S.~P.}\ \bibnamefont {Besedin}}, \bibinfo {author}
  {\bibfnamefont {M.~A.}\ \bibnamefont {Kuzovnikov}}, \bibinfo {author}
  {\bibfnamefont {S.}~\bibnamefont {Mozaffari}}, \bibinfo {author}
  {\bibfnamefont {L.}~\bibnamefont {Balicas}}, \bibinfo {author} {\bibfnamefont
  {F.~F.}\ \bibnamefont {Balakirev}}, \bibinfo {author} {\bibfnamefont {D.~E.}\
  \bibnamefont {Graf}}, \bibinfo {author} {\bibfnamefont {V.~B.}\ \bibnamefont
  {Prakapenka}}, \bibinfo {author} {\bibfnamefont {E.}~\bibnamefont
  {Greenberg}}, \bibinfo {author} {\bibfnamefont {D.~A.}\ \bibnamefont
  {Knyazev}}, \bibinfo {author} {\bibfnamefont {M.}~\bibnamefont {Tkacz}}, \
  and\ \bibinfo {author} {\bibfnamefont {M.~I.}\ \bibnamefont {Eremets}},\
  }\href {\doibase 10.1038/s41586-019-1201-8} {\bibfield  {journal} {\bibinfo
  {journal} {Nature}\ }\textbf {\bibinfo {volume} {569}},\ \bibinfo {pages}
  {528} (\bibinfo {year} {2019})}\BibitemShut {NoStop}%
\bibitem [{\citenamefont {Pepin}\ \emph {et~al.}(2017)\citenamefont {Pepin},
  \citenamefont {Geneste}, \citenamefont {Dewaele}, \citenamefont {Mezouar},\
  and\ \citenamefont {Loubeyre}}]{FeH5}%
  \BibitemOpen
  \bibfield  {author} {\bibinfo {author} {\bibfnamefont {C.~M.}\ \bibnamefont
  {Pepin}}, \bibinfo {author} {\bibfnamefont {G.}~\bibnamefont {Geneste}},
  \bibinfo {author} {\bibfnamefont {A.}~\bibnamefont {Dewaele}}, \bibinfo
  {author} {\bibfnamefont {M.}~\bibnamefont {Mezouar}}, \ and\ \bibinfo
  {author} {\bibfnamefont {P.}~\bibnamefont {Loubeyre}},\ }\href {\doibase
  10.1126/science.aan0961} {\bibfield  {journal} {\bibinfo  {journal}
  {Science}\ }\textbf {\bibinfo {volume} {357}},\ \bibinfo {pages} {382}
  (\bibinfo {year} {2017})}\BibitemShut {NoStop}%
\bibitem [{\citenamefont {Li}\ \emph {et~al.}(2019)\citenamefont {Li},
  \citenamefont {Huang}, \citenamefont {Duan}, \citenamefont {Pickard},
  \citenamefont {Zhou}, \citenamefont {Xie}, \citenamefont {Zhuang},
  \citenamefont {Huang}, \citenamefont {Zhou}, \citenamefont {Liu},\ and\
  \citenamefont {Cui}}]{CeH9}%
  \BibitemOpen
  \bibfield  {author} {\bibinfo {author} {\bibfnamefont {X.}~\bibnamefont
  {Li}}, \bibinfo {author} {\bibfnamefont {X.}~\bibnamefont {Huang}}, \bibinfo
  {author} {\bibfnamefont {D.}~\bibnamefont {Duan}}, \bibinfo {author}
  {\bibfnamefont {C.~J.}\ \bibnamefont {Pickard}}, \bibinfo {author}
  {\bibfnamefont {D.}~\bibnamefont {Zhou}}, \bibinfo {author} {\bibfnamefont
  {H.}~\bibnamefont {Xie}}, \bibinfo {author} {\bibfnamefont {Q.}~\bibnamefont
  {Zhuang}}, \bibinfo {author} {\bibfnamefont {Y.}~\bibnamefont {Huang}},
  \bibinfo {author} {\bibfnamefont {Q.}~\bibnamefont {Zhou}}, \bibinfo {author}
  {\bibfnamefont {B.}~\bibnamefont {Liu}}, \ and\ \bibinfo {author}
  {\bibfnamefont {T.}~\bibnamefont {Cui}},\ }\href {\doibase
  10.1038/s41467-019-11330-6} {\bibfield  {journal} {\bibinfo  {journal} {Nat.
  Commun.}\ }\textbf {\bibinfo {volume} {10}},\ \bibinfo {pages} {3461}
  (\bibinfo {year} {2019})}\BibitemShut {NoStop}%
\bibitem [{\citenamefont {Troyan}\ \emph {et~al.}(2019)\citenamefont {Troyan},
  \citenamefont {Semenok}, \citenamefont {Kvashnin}, \citenamefont {Ivanova},
  \citenamefont {Prakapenka}, \citenamefont {Greenberg}, \citenamefont
  {Gavriliuk}, \citenamefont {Lyubutin}, \citenamefont {Struzhkin},\ and\
  \citenamefont {Oganov}}]{YH6}%
  \BibitemOpen
  \bibfield  {author} {\bibinfo {author} {\bibfnamefont {I.}~\bibnamefont
  {Troyan}}, \bibinfo {author} {\bibfnamefont {D.}~\bibnamefont {Semenok}},
  \bibinfo {author} {\bibfnamefont {A.}~\bibnamefont {Kvashnin}}, \bibinfo
  {author} {\bibfnamefont {A.}~\bibnamefont {Ivanova}}, \bibinfo {author}
  {\bibfnamefont {V.}~\bibnamefont {Prakapenka}}, \bibinfo {author}
  {\bibfnamefont {E.}~\bibnamefont {Greenberg}}, \bibinfo {author}
  {\bibfnamefont {A.}~\bibnamefont {Gavriliuk}}, \bibinfo {author}
  {\bibfnamefont {I.}~\bibnamefont {Lyubutin}}, \bibinfo {author}
  {\bibfnamefont {V.}~\bibnamefont {Struzhkin}}, \ and\ \bibinfo {author}
  {\bibfnamefont {A.}~\bibnamefont {Oganov}},\ }\href {\doibase
  arxiv.org/abs/1908.01534} {\  (\bibinfo {year} {2019}),\
  arxiv.org/abs/1908.01534}\BibitemShut {NoStop}%
\bibitem [{\citenamefont {Geballe}\ \emph {et~al.}(2018)\citenamefont
  {Geballe}, \citenamefont {Liu}, \citenamefont {Mishra}, \citenamefont
  {Ahart}, \citenamefont {Somayazulu}, \citenamefont {Meng}, \citenamefont
  {Baldini},\ and\ \citenamefont {Hemley}}]{LaH10-angew}%
  \BibitemOpen
  \bibfield  {author} {\bibinfo {author} {\bibfnamefont {Z.~M.}\ \bibnamefont
  {Geballe}}, \bibinfo {author} {\bibfnamefont {H.}~\bibnamefont {Liu}},
  \bibinfo {author} {\bibfnamefont {A.~K.}\ \bibnamefont {Mishra}}, \bibinfo
  {author} {\bibfnamefont {M.}~\bibnamefont {Ahart}}, \bibinfo {author}
  {\bibfnamefont {M.}~\bibnamefont {Somayazulu}}, \bibinfo {author}
  {\bibfnamefont {Y.}~\bibnamefont {Meng}}, \bibinfo {author} {\bibfnamefont
  {M.}~\bibnamefont {Baldini}}, \ and\ \bibinfo {author} {\bibfnamefont
  {R.~J.}\ \bibnamefont {Hemley}},\ }\href {\doibase 10.1002/anie.201709970}
  {\bibfield  {journal} {\bibinfo  {journal} {Angew. Chem. Int. Ed.}\ }\textbf
  {\bibinfo {volume} {57}},\ \bibinfo {pages} {688} (\bibinfo {year}
  {2018})}\BibitemShut {NoStop}%
\bibitem [{\citenamefont {Kruglov}\ \emph {et~al.}(2020)\citenamefont
  {Kruglov}, \citenamefont {Semenok}, \citenamefont {Song}, \citenamefont
  {Szczesniak}, \citenamefont {Wrona}, \citenamefont {Akashi}, \citenamefont
  {Davari~Esfahani}, \citenamefont {Duan}, \citenamefont {Cui}, \citenamefont
  {Kvashnin},\ and\ \citenamefont {Oganov}}]{Kruglov}%
  \BibitemOpen
  \bibfield  {author} {\bibinfo {author} {\bibfnamefont {I.~A.}\ \bibnamefont
  {Kruglov}}, \bibinfo {author} {\bibfnamefont {D.~V.}\ \bibnamefont
  {Semenok}}, \bibinfo {author} {\bibfnamefont {H.}~\bibnamefont {Song}},
  \bibinfo {author} {\bibfnamefont {R.}~\bibnamefont {Szczesniak}}, \bibinfo
  {author} {\bibfnamefont {I.~A.}\ \bibnamefont {Wrona}}, \bibinfo {author}
  {\bibfnamefont {R.}~\bibnamefont {Akashi}}, \bibinfo {author} {\bibfnamefont
  {M.~M.}\ \bibnamefont {Davari~Esfahani}}, \bibinfo {author} {\bibfnamefont
  {D.}~\bibnamefont {Duan}}, \bibinfo {author} {\bibfnamefont {T.}~\bibnamefont
  {Cui}}, \bibinfo {author} {\bibfnamefont {A.~G.}\ \bibnamefont {Kvashnin}}, \
  and\ \bibinfo {author} {\bibfnamefont {A.~R.}\ \bibnamefont {Oganov}},\
  }\href {\doibase 10.1103/PhysRevB.101.024508} {\bibfield  {journal} {\bibinfo
   {journal} {Phys. Rev. B}\ }\textbf {\bibinfo {volume} {101}},\ \bibinfo
  {pages} {024508} (\bibinfo {year} {2020})}\BibitemShut {NoStop}%
\bibitem [{\citenamefont {Liu}\ \emph {et~al.}(2019)\citenamefont {Liu},
  \citenamefont {Wang}, \citenamefont {Yi}, \citenamefont {Kim}, \citenamefont
  {Kim},\ and\ \citenamefont {Cho}}]{LiuLiangliang}%
  \BibitemOpen
  \bibfield  {author} {\bibinfo {author} {\bibfnamefont {L.}~\bibnamefont
  {Liu}}, \bibinfo {author} {\bibfnamefont {C.}~\bibnamefont {Wang}}, \bibinfo
  {author} {\bibfnamefont {S.}~\bibnamefont {Yi}}, \bibinfo {author}
  {\bibfnamefont {K.~W.}\ \bibnamefont {Kim}}, \bibinfo {author} {\bibfnamefont
  {J.}~\bibnamefont {Kim}}, \ and\ \bibinfo {author} {\bibfnamefont {J.-H.}\
  \bibnamefont {Cho}},\ }\href {\doibase 10.1103/PhysRevB.99.140501} {\bibfield
   {journal} {\bibinfo  {journal} {Phys. Rev. B}\ }\textbf {\bibinfo {volume}
  {99}},\ \bibinfo {pages} {140501(R)} (\bibinfo {year} {2019})}\BibitemShut
  {NoStop}%
\bibitem [{\citenamefont {Serin}\ \emph {et~al.}(1950)\citenamefont {Serin},
  \citenamefont {Reynolds},\ and\ \citenamefont {Nesbitt}}]{Isotopecoeff}%
  \BibitemOpen
  \bibfield  {author} {\bibinfo {author} {\bibfnamefont {B.}~\bibnamefont
  {Serin}}, \bibinfo {author} {\bibfnamefont {C.~A.}\ \bibnamefont {Reynolds}},
  \ and\ \bibinfo {author} {\bibfnamefont {L.~B.}\ \bibnamefont {Nesbitt}},\
  }\href {\doibase 10.1103/PhysRev.78.813} {\bibfield  {journal} {\bibinfo
  {journal} {Phys. Rev.}\ }\textbf {\bibinfo {volume} {78}},\ \bibinfo {pages}
  {813} (\bibinfo {year} {1950})}\BibitemShut {NoStop}%
\bibitem [{\citenamefont {Giannozzi}\ \emph {et~al.}(2009)\citenamefont
  {Giannozzi}, \citenamefont {Baroni}, \citenamefont {Bonini}, \citenamefont
  {Calandra}, \citenamefont {Car}, \citenamefont {Cavazzoni}, \citenamefont
  {Ceresoli}, \citenamefont {Chiarotti}, \citenamefont {Cococcioni},
  \citenamefont {Dabo}, \citenamefont {Corso}, \citenamefont {de~Gironcoli},
  \citenamefont {Fabris}, \citenamefont {Fratesi}, \citenamefont {Gebauer},
  \citenamefont {Gerstmann}, \citenamefont {Gougoussis}, \citenamefont
  {Kokalj}, \citenamefont {Lazzeri}, \citenamefont {Martin-Samos},
  \citenamefont {Marzari}, \citenamefont {Mauri}, \citenamefont {Mazzarello},
  \citenamefont {Paolini}, \citenamefont {Pasquarello}, \citenamefont
  {Paulatto}, \citenamefont {Sbraccia}, \citenamefont {Scandolo}, \citenamefont
  {Sclauzero}, \citenamefont {Seitsonen}, \citenamefont {Smogunov},
  \citenamefont {Umari},\ and\ \citenamefont {Wentzcovitch}}]{QE}%
  \BibitemOpen
  \bibfield  {author} {\bibinfo {author} {\bibfnamefont {P.}~\bibnamefont
  {Giannozzi}}, \bibinfo {author} {\bibfnamefont {S.}~\bibnamefont {Baroni}},
  \bibinfo {author} {\bibfnamefont {N.}~\bibnamefont {Bonini}}, \bibinfo
  {author} {\bibfnamefont {M.}~\bibnamefont {Calandra}}, \bibinfo {author}
  {\bibfnamefont {R.}~\bibnamefont {Car}}, \bibinfo {author} {\bibfnamefont
  {C.}~\bibnamefont {Cavazzoni}}, \bibinfo {author} {\bibfnamefont
  {D.}~\bibnamefont {Ceresoli}}, \bibinfo {author} {\bibfnamefont {G.~L.}\
  \bibnamefont {Chiarotti}}, \bibinfo {author} {\bibfnamefont {M.}~\bibnamefont
  {Cococcioni}}, \bibinfo {author} {\bibfnamefont {I.}~\bibnamefont {Dabo}},
  \bibinfo {author} {\bibfnamefont {A.~D.}\ \bibnamefont {Corso}}, \bibinfo
  {author} {\bibfnamefont {S.}~\bibnamefont {de~Gironcoli}}, \bibinfo {author}
  {\bibfnamefont {S.}~\bibnamefont {Fabris}}, \bibinfo {author} {\bibfnamefont
  {G.}~\bibnamefont {Fratesi}}, \bibinfo {author} {\bibfnamefont
  {R.}~\bibnamefont {Gebauer}}, \bibinfo {author} {\bibfnamefont
  {U.}~\bibnamefont {Gerstmann}}, \bibinfo {author} {\bibfnamefont
  {C.}~\bibnamefont {Gougoussis}}, \bibinfo {author} {\bibfnamefont
  {A.}~\bibnamefont {Kokalj}}, \bibinfo {author} {\bibfnamefont
  {M.}~\bibnamefont {Lazzeri}}, \bibinfo {author} {\bibfnamefont
  {L.}~\bibnamefont {Martin-Samos}}, \bibinfo {author} {\bibfnamefont
  {N.}~\bibnamefont {Marzari}}, \bibinfo {author} {\bibfnamefont
  {F.}~\bibnamefont {Mauri}}, \bibinfo {author} {\bibfnamefont
  {R.}~\bibnamefont {Mazzarello}}, \bibinfo {author} {\bibfnamefont
  {S.}~\bibnamefont {Paolini}}, \bibinfo {author} {\bibfnamefont
  {A.}~\bibnamefont {Pasquarello}}, \bibinfo {author} {\bibfnamefont
  {L.}~\bibnamefont {Paulatto}}, \bibinfo {author} {\bibfnamefont
  {C.}~\bibnamefont {Sbraccia}}, \bibinfo {author} {\bibfnamefont
  {S.}~\bibnamefont {Scandolo}}, \bibinfo {author} {\bibfnamefont
  {G.}~\bibnamefont {Sclauzero}}, \bibinfo {author} {\bibfnamefont {A.~P.}\
  \bibnamefont {Seitsonen}}, \bibinfo {author} {\bibfnamefont {A.}~\bibnamefont
  {Smogunov}}, \bibinfo {author} {\bibfnamefont {P.}~\bibnamefont {Umari}}, \
  and\ \bibinfo {author} {\bibfnamefont {R.~M.}\ \bibnamefont {Wentzcovitch}},\
  }\href {http://stacks.iop.org/0953-8984/21/i=39/a=395502} {\bibfield
  {journal} {\bibinfo  {journal} {J. Phys. Condens. Matter}\ }\textbf {\bibinfo
  {volume} {21}},\ \bibinfo {pages} {395502} (\bibinfo {year}
  {2009})}\BibitemShut {NoStop}%
\bibitem [{\citenamefont {Giannozzi}\ \emph {et~al.}(2017)\citenamefont
  {Giannozzi}, \citenamefont {Andreussi}, \citenamefont {Brumme}, \citenamefont
  {Bunau}, \citenamefont {Nardelli}, \citenamefont {Calandra}, \citenamefont
  {Car}, \citenamefont {Cavazzoni}, \citenamefont {Ceresoli}, \citenamefont
  {Cococcioni}, \citenamefont {Colonna}, \citenamefont {Carnimeo},
  \citenamefont {Corso}, \citenamefont {de~Gironcoli}, \citenamefont {Delugas},
  \citenamefont {Jr}, \citenamefont {Ferretti}, \citenamefont {Floris},
  \citenamefont {Fratesi}, \citenamefont {Fugallo}, \citenamefont {Gebauer},
  \citenamefont {Gerstmann}, \citenamefont {Giustino}, \citenamefont {Gorni},
  \citenamefont {Jia}, \citenamefont {Kawamura}, \citenamefont {Ko},
  \citenamefont {Kokalj}, \citenamefont {Küçükbenli}, \citenamefont
  {Lazzeri}, \citenamefont {Marsili}, \citenamefont {Marzari}, \citenamefont
  {Mauri}, \citenamefont {Nguyen}, \citenamefont {Nguyen}, \citenamefont {de-la
  Roza}, \citenamefont {Paulatto}, \citenamefont {Poncé}, \citenamefont
  {Rocca}, \citenamefont {Sabatini}, \citenamefont {Santra}, \citenamefont
  {Schlipf}, \citenamefont {Seitsonen}, \citenamefont {Smogunov}, \citenamefont
  {Timrov}, \citenamefont {Thonhauser}, \citenamefont {Umari}, \citenamefont
  {Vast}, \citenamefont {Wu},\ and\ \citenamefont {Baroni}}]{QE2}%
  \BibitemOpen
  \bibfield  {author} {\bibinfo {author} {\bibfnamefont {P.}~\bibnamefont
  {Giannozzi}}, \bibinfo {author} {\bibfnamefont {O.}~\bibnamefont
  {Andreussi}}, \bibinfo {author} {\bibfnamefont {T.}~\bibnamefont {Brumme}},
  \bibinfo {author} {\bibfnamefont {O.}~\bibnamefont {Bunau}}, \bibinfo
  {author} {\bibfnamefont {M.~B.}\ \bibnamefont {Nardelli}}, \bibinfo {author}
  {\bibfnamefont {M.}~\bibnamefont {Calandra}}, \bibinfo {author}
  {\bibfnamefont {R.}~\bibnamefont {Car}}, \bibinfo {author} {\bibfnamefont
  {C.}~\bibnamefont {Cavazzoni}}, \bibinfo {author} {\bibfnamefont
  {D.}~\bibnamefont {Ceresoli}}, \bibinfo {author} {\bibfnamefont
  {M.}~\bibnamefont {Cococcioni}}, \bibinfo {author} {\bibfnamefont
  {N.}~\bibnamefont {Colonna}}, \bibinfo {author} {\bibfnamefont
  {I.}~\bibnamefont {Carnimeo}}, \bibinfo {author} {\bibfnamefont {A.~D.}\
  \bibnamefont {Corso}}, \bibinfo {author} {\bibfnamefont {S.}~\bibnamefont
  {de~Gironcoli}}, \bibinfo {author} {\bibfnamefont {P.}~\bibnamefont
  {Delugas}}, \bibinfo {author} {\bibfnamefont {R.~A.~D.}\ \bibnamefont {Jr}},
  \bibinfo {author} {\bibfnamefont {A.}~\bibnamefont {Ferretti}}, \bibinfo
  {author} {\bibfnamefont {A.}~\bibnamefont {Floris}}, \bibinfo {author}
  {\bibfnamefont {G.}~\bibnamefont {Fratesi}}, \bibinfo {author} {\bibfnamefont
  {G.}~\bibnamefont {Fugallo}}, \bibinfo {author} {\bibfnamefont
  {R.}~\bibnamefont {Gebauer}}, \bibinfo {author} {\bibfnamefont
  {U.}~\bibnamefont {Gerstmann}}, \bibinfo {author} {\bibfnamefont
  {F.}~\bibnamefont {Giustino}}, \bibinfo {author} {\bibfnamefont
  {T.}~\bibnamefont {Gorni}}, \bibinfo {author} {\bibfnamefont
  {J.}~\bibnamefont {Jia}}, \bibinfo {author} {\bibfnamefont {M.}~\bibnamefont
  {Kawamura}}, \bibinfo {author} {\bibfnamefont {H.-Y.}\ \bibnamefont {Ko}},
  \bibinfo {author} {\bibfnamefont {A.}~\bibnamefont {Kokalj}}, \bibinfo
  {author} {\bibfnamefont {E.}~\bibnamefont {Küçükbenli}}, \bibinfo {author}
  {\bibfnamefont {M.}~\bibnamefont {Lazzeri}}, \bibinfo {author} {\bibfnamefont
  {M.}~\bibnamefont {Marsili}}, \bibinfo {author} {\bibfnamefont
  {N.}~\bibnamefont {Marzari}}, \bibinfo {author} {\bibfnamefont
  {F.}~\bibnamefont {Mauri}}, \bibinfo {author} {\bibfnamefont {N.~L.}\
  \bibnamefont {Nguyen}}, \bibinfo {author} {\bibfnamefont {H.-V.}\
  \bibnamefont {Nguyen}}, \bibinfo {author} {\bibfnamefont {A.~O.}\
  \bibnamefont {de-la Roza}}, \bibinfo {author} {\bibfnamefont
  {L.}~\bibnamefont {Paulatto}}, \bibinfo {author} {\bibfnamefont
  {S.}~\bibnamefont {Poncé}}, \bibinfo {author} {\bibfnamefont
  {D.}~\bibnamefont {Rocca}}, \bibinfo {author} {\bibfnamefont
  {R.}~\bibnamefont {Sabatini}}, \bibinfo {author} {\bibfnamefont
  {B.}~\bibnamefont {Santra}}, \bibinfo {author} {\bibfnamefont
  {M.}~\bibnamefont {Schlipf}}, \bibinfo {author} {\bibfnamefont {A.~P.}\
  \bibnamefont {Seitsonen}}, \bibinfo {author} {\bibfnamefont {A.}~\bibnamefont
  {Smogunov}}, \bibinfo {author} {\bibfnamefont {I.}~\bibnamefont {Timrov}},
  \bibinfo {author} {\bibfnamefont {T.}~\bibnamefont {Thonhauser}}, \bibinfo
  {author} {\bibfnamefont {P.}~\bibnamefont {Umari}}, \bibinfo {author}
  {\bibfnamefont {N.}~\bibnamefont {Vast}}, \bibinfo {author} {\bibfnamefont
  {X.}~\bibnamefont {Wu}}, \ and\ \bibinfo {author} {\bibfnamefont
  {S.}~\bibnamefont {Baroni}},\ }\href
  {http://stacks.iop.org/0953-8984/29/i=46/a=465901} {\bibfield  {journal}
  {\bibinfo  {journal} {J. Phys. Condens. Matter}\ }\textbf {\bibinfo {volume}
  {29}},\ \bibinfo {pages} {465901} (\bibinfo {year} {2017})}\BibitemShut
  {NoStop}%
\bibitem [{\citenamefont {Billeter}\ \emph {et~al.}(2003)\citenamefont
  {Billeter}, \citenamefont {Curioni},\ and\ \citenamefont {Andreoni}}]{bfgs}%
  \BibitemOpen
  \bibfield  {author} {\bibinfo {author} {\bibfnamefont {S.~R.}\ \bibnamefont
  {Billeter}}, \bibinfo {author} {\bibfnamefont {A.}~\bibnamefont {Curioni}}, \
  and\ \bibinfo {author} {\bibfnamefont {W.}~\bibnamefont {Andreoni}},\ }\href
  {\doibase https://doi.org/10.1016/S0927-0256(03)00043-0} {\bibfield
  {journal} {\bibinfo  {journal} {Comput. Mater. Sci.}\ }\textbf {\bibinfo
  {volume} {27}},\ \bibinfo {pages} {437 } (\bibinfo {year}
  {2003})}\BibitemShut {NoStop}%
\bibitem [{\citenamefont {Baroni}\ \emph {et~al.}(2001)\citenamefont {Baroni},
  \citenamefont {de~Gironcoli}, \citenamefont {Dal~Corso},\ and\ \citenamefont
  {Giannozzi}}]{BaroniDFPT}%
  \BibitemOpen
  \bibfield  {author} {\bibinfo {author} {\bibfnamefont {S.}~\bibnamefont
  {Baroni}}, \bibinfo {author} {\bibfnamefont {S.}~\bibnamefont
  {de~Gironcoli}}, \bibinfo {author} {\bibfnamefont {A.}~\bibnamefont
  {Dal~Corso}}, \ and\ \bibinfo {author} {\bibfnamefont {P.}~\bibnamefont
  {Giannozzi}},\ }\href {\doibase 10.1103/RevModPhys.73.515} {\bibfield
  {journal} {\bibinfo  {journal} {Rev. Mod. Phys.}\ }\textbf {\bibinfo {volume}
  {73}},\ \bibinfo {pages} {515} (\bibinfo {year} {2001})}\BibitemShut
  {NoStop}%
\bibitem [{\citenamefont {Eliashberg}(1960)}]{Eliashberg1960A}%
  \BibitemOpen
  \bibfield  {author} {\bibinfo {author} {\bibfnamefont {G.~M.}\ \bibnamefont
  {Eliashberg}},\ }\href {http://www.jetp.ac.ru/cgi-bin/dn/e_011_03_0696.pdf}
  {\bibfield  {journal} {\bibinfo  {journal} {J. Exp. Theor. Phys.}\ }\textbf
  {\bibinfo {volume} {11}},\ \bibinfo {pages} {696} (\bibinfo {year}
  {1960})}\BibitemShut {NoStop}%
\bibitem [{\citenamefont {Marsiglio}\ \emph {et~al.}(1988)\citenamefont
  {Marsiglio}, \citenamefont {Schossmann},\ and\ \citenamefont
  {Carbotte}}]{Marsiglio1988A}%
  \BibitemOpen
  \bibfield  {author} {\bibinfo {author} {\bibfnamefont {F.}~\bibnamefont
  {Marsiglio}}, \bibinfo {author} {\bibfnamefont {M.}~\bibnamefont
  {Schossmann}}, \ and\ \bibinfo {author} {\bibfnamefont {J.~P.}\ \bibnamefont
  {Carbotte}},\ }\href {\doibase 10.1103/PhysRevB.37.4965} {\bibfield
  {journal} {\bibinfo  {journal} {Phys. Rev. B}\ }\textbf {\bibinfo {volume}
  {37}},\ \bibinfo {pages} {4965} (\bibinfo {year} {1988})}\BibitemShut
  {NoStop}%
\bibitem [{\citenamefont {Durajski}(2016)}]{Durajski-50-2016}%
  \BibitemOpen
  \bibfield  {author} {\bibinfo {author} {\bibfnamefont {A.~P.}\ \bibnamefont
  {Durajski}},\ }\href {\doibase 10.1038/srep38570} {\bibfield  {journal}
  {\bibinfo  {journal} {Sci. Rep.}\ }\textbf {\bibinfo {volume} {6}},\ \bibinfo
  {pages} {38570} (\bibinfo {year} {2016})}\BibitemShut {NoStop}%
\bibitem [{\citenamefont {Szczesniak}\ and\ \citenamefont
  {Zemla}(2015)}]{DominZemla}%
  \BibitemOpen
  \bibfield  {author} {\bibinfo {author} {\bibfnamefont {D.}~\bibnamefont
  {Szczesniak}}\ and\ \bibinfo {author} {\bibfnamefont {T.~P.}\ \bibnamefont
  {Zemla}},\ }\href {http://stacks.iop.org/0953-2048/28/i=8/a=085018}
  {\bibfield  {journal} {\bibinfo  {journal} {Supercond. Sci. Technol.}\
  }\textbf {\bibinfo {volume} {28}},\ \bibinfo {pages} {085018} (\bibinfo
  {year} {2015})}\BibitemShut {NoStop}%
\bibitem [{\citenamefont {Szczesniak}(2006)}]{szczesniak2006A}%
  \BibitemOpen
  \bibfield  {author} {\bibinfo {author} {\bibfnamefont {R.}~\bibnamefont
  {Szczesniak}},\ }\href {\doibase 10.12693/APhysPolA.109.179} {\bibfield
  {journal} {\bibinfo  {journal} {Acta Phys. Pol. A}\ }\textbf {\bibinfo
  {volume} {109}},\ \bibinfo {pages} {179} (\bibinfo {year}
  {2006})}\BibitemShut {NoStop}%
\bibitem [{\citenamefont {Durajski}\ and\ \citenamefont
  {Szczesniak}(2018{\natexlab{a}})}]{Durajski-64-2018}%
  \BibitemOpen
  \bibfield  {author} {\bibinfo {author} {\bibfnamefont {A.~P.}\ \bibnamefont
  {Durajski}}\ and\ \bibinfo {author} {\bibfnamefont {R.}~\bibnamefont
  {Szczesniak}},\ }\href {\doibase 10.1063/1.5031202} {\bibfield  {journal}
  {\bibinfo  {journal} {J. Chem. Phys.}\ }\textbf {\bibinfo {volume} {149}},\
  \bibinfo {pages} {074101} (\bibinfo {year} {2018}{\natexlab{a}})}\BibitemShut
  {NoStop}%
\bibitem [{\citenamefont {Wiendlocha}\ \emph {et~al.}(2016)\citenamefont
  {Wiendlocha}, \citenamefont {Szczesniak}, \citenamefont {Durajski},\ and\
  \citenamefont {Muras}}]{Wiendlocha}%
  \BibitemOpen
  \bibfield  {author} {\bibinfo {author} {\bibfnamefont {B.}~\bibnamefont
  {Wiendlocha}}, \bibinfo {author} {\bibfnamefont {R.}~\bibnamefont
  {Szczesniak}}, \bibinfo {author} {\bibfnamefont {A.~P.}\ \bibnamefont
  {Durajski}}, \ and\ \bibinfo {author} {\bibfnamefont {M.}~\bibnamefont
  {Muras}},\ }\href {\doibase 10.1103/PhysRevB.94.134517} {\bibfield  {journal}
  {\bibinfo  {journal} {Phys. Rev. B}\ }\textbf {\bibinfo {volume} {94}},\
  \bibinfo {pages} {134517} (\bibinfo {year} {2016})}\BibitemShut {NoStop}%
\bibitem [{\citenamefont {Carbotte}\ \emph {et~al.}(2019)\citenamefont
  {Carbotte}, \citenamefont {Nicol},\ and\ \citenamefont
  {Timusk}}]{PhysRevB.100.094505}%
  \BibitemOpen
  \bibfield  {author} {\bibinfo {author} {\bibfnamefont {J.~P.}\ \bibnamefont
  {Carbotte}}, \bibinfo {author} {\bibfnamefont {E.~J.}\ \bibnamefont {Nicol}},
  \ and\ \bibinfo {author} {\bibfnamefont {T.}~\bibnamefont {Timusk}},\ }\href
  {\doibase 10.1103/PhysRevB.100.094505} {\bibfield  {journal} {\bibinfo
  {journal} {Phys. Rev. B}\ }\textbf {\bibinfo {volume} {100}},\ \bibinfo
  {pages} {094505} (\bibinfo {year} {2019})}\BibitemShut {NoStop}%
\bibitem [{SM()}]{SM}%
  \BibitemOpen
  \href@noop {} {}\bibinfo {note} {See Supplemental Material at
  \url{http://link.aps.org/supplemental/XXXXXXXX} for further plots of
  calculated phonon spectrum of the fcc ${\mathrm{LaH}}_{10}$ phase at 210 GPa
  and superconducting critical temperatures calculated using Migdal-Eliashberg
  equations with anharmonic phonons.}\BibitemShut {Stop}%
\bibitem [{\citenamefont {Wang}\ \emph {et~al.}(2019)\citenamefont {Wang},
  \citenamefont {Yi},\ and\ \citenamefont {Cho}}]{chongze-prb}%
  \BibitemOpen
  \bibfield  {author} {\bibinfo {author} {\bibfnamefont {C.}~\bibnamefont
  {Wang}}, \bibinfo {author} {\bibfnamefont {S.}~\bibnamefont {Yi}}, \ and\
  \bibinfo {author} {\bibfnamefont {J.-H.}\ \bibnamefont {Cho}},\ }\href
  {\doibase 10.1103/PhysRevB.100.060502} {\bibfield  {journal} {\bibinfo
  {journal} {Phys. Rev. B}\ }\textbf {\bibinfo {volume} {100}},\ \bibinfo
  {pages} {060502(R)} (\bibinfo {year} {2019})}\BibitemShut {NoStop}%
\bibitem [{\citenamefont {Jarlborg}\ and\ \citenamefont
  {Bianconi}(2016)}]{BianconiSciRep}%
  \BibitemOpen
  \bibfield  {author} {\bibinfo {author} {\bibfnamefont {T.}~\bibnamefont
  {Jarlborg}}\ and\ \bibinfo {author} {\bibfnamefont {A.}~\bibnamefont
  {Bianconi}},\ }\href {\doibase 10.1038/srep24816} {\bibfield  {journal}
  {\bibinfo  {journal} {Sci. Rep.}\ }\textbf {\bibinfo {volume} {6}},\ \bibinfo
  {pages} {24816} (\bibinfo {year} {2016})}\BibitemShut {NoStop}%
\bibitem [{\citenamefont {Quan}\ and\ \citenamefont
  {Pickett}(2016)}]{PhysRevB.93.104526}%
  \BibitemOpen
  \bibfield  {author} {\bibinfo {author} {\bibfnamefont {Y.}~\bibnamefont
  {Quan}}\ and\ \bibinfo {author} {\bibfnamefont {W.~E.}\ \bibnamefont
  {Pickett}},\ }\href {\doibase 10.1103/PhysRevB.93.104526} {\bibfield
  {journal} {\bibinfo  {journal} {Phys. Rev. B}\ }\textbf {\bibinfo {volume}
  {93}},\ \bibinfo {pages} {104526} (\bibinfo {year} {2016})}\BibitemShut
  {NoStop}%
\bibitem [{\citenamefont {Krzyzosiak}\ \emph {et~al.}(2019)\citenamefont
  {Krzyzosiak}, \citenamefont {Gonczarek}, \citenamefont {Gonczarek},\ and\
  \citenamefont {L.}}]{Gonczarek2}%
  \BibitemOpen
  \bibfield  {author} {\bibinfo {author} {\bibfnamefont {M.}~\bibnamefont
  {Krzyzosiak}}, \bibinfo {author} {\bibfnamefont {R.}~\bibnamefont
  {Gonczarek}}, \bibinfo {author} {\bibfnamefont {A.}~\bibnamefont
  {Gonczarek}}, \ and\ \bibinfo {author} {\bibfnamefont {J.}~\bibnamefont
  {L.}},\ }\href {\doibase 10.1038/s41598-018-36733-1} {\bibfield  {journal}
  {\bibinfo  {journal} {Sci. Rep.}\ }\textbf {\bibinfo {volume} {9}},\ \bibinfo
  {pages} {2181} (\bibinfo {year} {2019})}\BibitemShut {NoStop}%
\bibitem [{\citenamefont {Durajski}\ and\ \citenamefont
  {Szczesniak}(2017)}]{Durajski-54-2017}%
  \BibitemOpen
  \bibfield  {author} {\bibinfo {author} {\bibfnamefont {A.~P.}\ \bibnamefont
  {Durajski}}\ and\ \bibinfo {author} {\bibfnamefont {R.}~\bibnamefont
  {Szczesniak}},\ }\href {\doibase 10.1038/s41598-017-04714-5} {\bibfield
  {journal} {\bibinfo  {journal} {Sci. Rep.}\ }\textbf {\bibinfo {volume}
  {7}},\ \bibinfo {pages} {4473} (\bibinfo {year} {2017})}\BibitemShut
  {NoStop}%
\bibitem [{\citenamefont {Durajski}\ and\ \citenamefont
  {Szczesniak}(2018{\natexlab{b}})}]{Durajski-65-2018}%
  \BibitemOpen
  \bibfield  {author} {\bibinfo {author} {\bibfnamefont {A.~P.}\ \bibnamefont
  {Durajski}}\ and\ \bibinfo {author} {\bibfnamefont {R.}~\bibnamefont
  {Szczesniak}},\ }\href {\doibase https://doi.org/10.1016/j.physc.2018.09.004}
  {\bibfield  {journal} {\bibinfo  {journal} {Physica C}\ }\textbf {\bibinfo
  {volume} {554}},\ \bibinfo {pages} {38 } (\bibinfo {year}
  {2018}{\natexlab{b}})}\BibitemShut {NoStop}%
\bibitem [{\citenamefont {Errea}\ \emph {et~al.}(2020)\citenamefont {Errea},
  \citenamefont {Belli}, \citenamefont {Monacelli}, \citenamefont {Sanna},
  \citenamefont {Koretsune}, \citenamefont {Tadano}, \citenamefont {Bianco},
  \citenamefont {Calandra}, \citenamefont {Arita}, \citenamefont {Mauri},\ and\
  \citenamefont {Flores-Livas}}]{Errea2020}%
  \BibitemOpen
  \bibfield  {author} {\bibinfo {author} {\bibfnamefont {I.}~\bibnamefont
  {Errea}}, \bibinfo {author} {\bibfnamefont {F.}~\bibnamefont {Belli}},
  \bibinfo {author} {\bibfnamefont {L.}~\bibnamefont {Monacelli}}, \bibinfo
  {author} {\bibfnamefont {A.}~\bibnamefont {Sanna}}, \bibinfo {author}
  {\bibfnamefont {T.}~\bibnamefont {Koretsune}}, \bibinfo {author}
  {\bibfnamefont {T.}~\bibnamefont {Tadano}}, \bibinfo {author} {\bibfnamefont
  {R.}~\bibnamefont {Bianco}}, \bibinfo {author} {\bibfnamefont
  {M.}~\bibnamefont {Calandra}}, \bibinfo {author} {\bibfnamefont
  {R.}~\bibnamefont {Arita}}, \bibinfo {author} {\bibfnamefont
  {F.}~\bibnamefont {Mauri}}, \ and\ \bibinfo {author} {\bibfnamefont {J.~A.}\
  \bibnamefont {Flores-Livas}},\ }\href {\doibase
  doi.org/10.1038/s41586-020-1955-z} {\bibfield  {journal} {\bibinfo  {journal}
  {Nature}\ }\textbf {\bibinfo {volume} {578}},\ \bibinfo {pages} {66}
  (\bibinfo {year} {2020})}\BibitemShut {NoStop}%
\bibitem [{\citenamefont {Akashi}\ \emph {et~al.}(2015)\citenamefont {Akashi},
  \citenamefont {Kawamura}, \citenamefont {Tsuneyuki}, \citenamefont {Nomura},\
  and\ \citenamefont {Arita}}]{AkashiH2S}%
  \BibitemOpen
  \bibfield  {author} {\bibinfo {author} {\bibfnamefont {R.}~\bibnamefont
  {Akashi}}, \bibinfo {author} {\bibfnamefont {M.}~\bibnamefont {Kawamura}},
  \bibinfo {author} {\bibfnamefont {S.}~\bibnamefont {Tsuneyuki}}, \bibinfo
  {author} {\bibfnamefont {Y.}~\bibnamefont {Nomura}}, \ and\ \bibinfo {author}
  {\bibfnamefont {R.}~\bibnamefont {Arita}},\ }\href {\doibase
  10.1103/PhysRevB.91.224513} {\bibfield  {journal} {\bibinfo  {journal} {Phys.
  Rev. B}\ }\textbf {\bibinfo {volume} {91}},\ \bibinfo {pages} {224513}
  (\bibinfo {year} {2015})}\BibitemShut {NoStop}%
\bibitem [{\citenamefont {Errea}\ \emph {et~al.}(2015)\citenamefont {Errea},
  \citenamefont {Calandra}, \citenamefont {Pickard}, \citenamefont {Nelson},
  \citenamefont {Needs}, \citenamefont {Li}, \citenamefont {Liu}, \citenamefont
  {Zhang}, \citenamefont {Ma},\ and\ \citenamefont {Mauri}}]{Errea2015A}%
  \BibitemOpen
  \bibfield  {author} {\bibinfo {author} {\bibfnamefont {I.}~\bibnamefont
  {Errea}}, \bibinfo {author} {\bibfnamefont {M.}~\bibnamefont {Calandra}},
  \bibinfo {author} {\bibfnamefont {C.~J.}\ \bibnamefont {Pickard}}, \bibinfo
  {author} {\bibfnamefont {J.}~\bibnamefont {Nelson}}, \bibinfo {author}
  {\bibfnamefont {R.~J.}\ \bibnamefont {Needs}}, \bibinfo {author}
  {\bibfnamefont {Y.}~\bibnamefont {Li}}, \bibinfo {author} {\bibfnamefont
  {H.}~\bibnamefont {Liu}}, \bibinfo {author} {\bibfnamefont {Y.}~\bibnamefont
  {Zhang}}, \bibinfo {author} {\bibfnamefont {Y.}~\bibnamefont {Ma}}, \ and\
  \bibinfo {author} {\bibfnamefont {F.}~\bibnamefont {Mauri}},\ }\href
  {\doibase 10.1103/PhysRevLett.114.157004} {\bibfield  {journal} {\bibinfo
  {journal} {Phys. Rev. Lett.}\ }\textbf {\bibinfo {volume} {114}},\ \bibinfo
  {pages} {157004} (\bibinfo {year} {2015})}\BibitemShut {NoStop}%
\bibitem [{\citenamefont {Mendez-Moreno}(2019)}]{Mendez-Moreno}%
  \BibitemOpen
  \bibfield  {author} {\bibinfo {author} {\bibfnamefont {R.~M.}\ \bibnamefont
  {Mendez-Moreno}},\ }\href {\doibase 10.1155/2019/6795250} {\bibfield
  {journal} {\bibinfo  {journal} {Adv. Cond. Matter Phys.}\ }\textbf {\bibinfo
  {volume} {2019}},\ \bibinfo {pages} {6795250} (\bibinfo {year}
  {2019})}\BibitemShut {NoStop}%
\bibitem [{\citenamefont {Talantsev}(2019)}]{Talantsev_2019}%
  \BibitemOpen
  \bibfield  {author} {\bibinfo {author} {\bibfnamefont {E.~F.}\ \bibnamefont
  {Talantsev}},\ }\href {\doibase 10.1088/2053-1591/ab3bbb} {\bibfield
  {journal} {\bibinfo  {journal} {Mater. Res. Express}\ }\textbf {\bibinfo
  {volume} {6}},\ \bibinfo {pages} {106002} (\bibinfo {year}
  {2019})}\BibitemShut {NoStop}%
\bibitem [{\citenamefont {Szczesniak}\ and\ \citenamefont
  {Durajski}(2017)}]{SZCZESNIAK201730}%
  \BibitemOpen
  \bibfield  {author} {\bibinfo {author} {\bibfnamefont {R.}~\bibnamefont
  {Szczesniak}}\ and\ \bibinfo {author} {\bibfnamefont {A.}~\bibnamefont
  {Durajski}},\ }\href {\doibase https://doi.org/10.1016/j.ssc.2016.10.012}
  {\bibfield  {journal} {\bibinfo  {journal} {Solid State Commun.}\ }\textbf
  {\bibinfo {volume} {249}},\ \bibinfo {pages} {30 } (\bibinfo {year}
  {2017})}\BibitemShut {NoStop}%
\end{thebibliography}%

%
\end{document}